\documentclass[fleqn,usenatbib]{mnras}
\usepackage{newtxtext,newtxmath}
\usepackage[T1]{fontenc}
\DeclareRobustCommand{\VAN}[3]{#2}
\let\VANthebibliography\thebibliography
\def\thebibliography{\DeclareRobustCommand{\VAN}[3]{##3}\VANthebibliography} 

\usepackage{graphicx}	
\usepackage{amsmath}	
\usepackage{multirow}
\usepackage{xcolor}
\usepackage{orcidlink}
\setcitestyle{notesep={ }}
\usepackage{float}
\usepackage[caption = false]{subfig}
\usepackage{float}

\title[Utilizing Stellar Mass Estimates to Identify Gravitational Wave Host Galaxies]{Utilizing Stellar Mass Estimates to Identify Gravitational Wave Host Galaxies}

\author[M. Pálfi, G. Dálya and P. Raffai]{
M. Pálfi,$^{1}$ \orcidlink{0000-0001-5942-0470} \thanks{E-mail: marika97@student.elte.hu}
G. D\'alya$^{2,3}$  \orcidlink{0000-0003-3258-5763}
and P. Raffai$^{1,4} \orcidlink{0000-0001-7576-0141}$
\\
$^{1}$ 
Institute of Physics and Astronomy, ELTE E\"otv\"os Lor\'and University, 1117 Budapest, Hungary \\
$^{2}${L2IT, Laboratoire des 2 Infinis - Toulouse, Universit\'e de Toulouse, CNRS/IN2P3, UPS, F-31062 Toulouse Cedex 9, France} \\
$^{3}${Department of Physics and Astronomy, Universiteit Gent, B-9000 Ghent, Belgium} \\
$^{4}$HUN-REN–ELTE Extragalactic Astrophysics Research Group, 1117 Budapest, Hungary
}

\date{Accepted XXX. Received YYY; in original form ZZZ}

\pubyear{\the\year{}}

\begin{document}
\label{firstpage}
\pagerange{\pageref{firstpage}--\pageref{lastpage}}
\maketitle

\begin{abstract}
 Stellar mass can enhance the ranking of potential hosts for compact binary coalescences identified by ground-based gravitational-wave detectors within large localisation areas containing even thousands of galaxies. Despite its benefits, accurate stellar mass estimation is often time-consuming and computationally intensive. In this study, we implement four stellar mass estimation methods based on infrared magnitudes and compare them with values estimated with spectral energy distribution fitting from GAMA DR3, revealing strong correlations. We also introduce a method to calibrate the results from these estimation methods to match the reference values. Our analysis of simulated binary black hole events demonstrates that incorporating stellar mass improves the rank of actual hosts $\sim$80 per cent of cases. Furthermore, the improvement is comparable when stellar masses are derived from the tested estimation methods to when they are obtained directly from the simulated galaxy catalogue, demonstrating that simple stellar mass estimates can provide a computationally efficient alternative.
\end{abstract}


\section{Introduction}
Combining gravitational wave \citep[GW;][]{Barry_Weiss} and electromagnetic observations provides an opportunity to better understand astrophysical events, determine their origins, measure cosmological parameters, and probe their environments \citep{bright_siren, Abbott_2017}.
The large number of galaxies in the localisation volume of a compact binary coalescence (CBC)  event \citep[see table$~$1 of][]{O3cos} complicates the identification of the host galaxy.
\citet{Li_Williams} assumed that CBCs are more likely to occur in galaxies with larger $M_\ast$. They applied three stellar mass estimation (SME) methods \citep[one based on \textit{B} band magnitude, one using spectral energy distribution (SED) fitting method and one from][]{Cluver_2014} to the galaxies in the  Gravitational Wave Galaxy Catalogue \citep*[GWGC; see][]{White_2011} and the Galaxy List for the Advanced Detector Era \citep[GLADE; see][]{GLADE} catalogues, and they compared these to the very accurate stellar masses of the “Stripe 82” region \citep{Bundy_2015}.
\citet{Artale_1} confirmed the assumption of \citet{Li_Williams} by showing in their simulation that the $M_\ast$ strongly correlates with the merger rate per galaxy in the local universe ($z\leq0.1$). They also observed weaker correlations between the merger rate per galaxy and both the star formation rate (SFR) and metallicity, as these correlations exhibit larger scatter compared to the correlation between the merger rate per galaxy and $M_\ast$.
\citet{Artale_2} found that the strong correlation between the merger rate per galaxy and the $M_\ast$ holds for all studied redshift ranges ($z \in [0, 0.1]$, [0.93, 1.13], [1.87, 2.12], and [5.73, 6.51]). 
Building on these results, \citet{Ducoin} estimated the stellar mass of galaxies in the GLADE catalogue using the method of \citet{Kettlety_18} and developed a ranking process that considers the stellar mass. They tested this approach on GW170817, the only CBC event with an electromagnetic counterpart (conventionally referred to as a bright siren), and found that including the stellar mass improves the rank of the confirmed host galaxy. \citet{Artale_2020} found that the correlation between the host galaxy probability and the $M_\ast$ is not simply linear and proposed a ranking method based on simulated host galaxy probabilities.
Stellar mass can also help assess the probability of galaxies hosting a dark siren, a CBC event without a detected electromagnetic counterpart. Recently, \citet{Hanselman} examined how incorrectly assumed weighting schemes, including those based on stellar mass, affect Hubble constant estimates with GWs \citep[see, e.g.,][]{O3cos}. 

The stellar mass ($M_\ast$) is the total mass of stars and stellar remnants in a galaxy, excluding the mass of interstellar gas, dust, and dark matter \citep[see, e.g.,][]{Courteau_2014}.  $M_\ast$ plays a crucial role in the formation and evolution of galaxies, as highlighted by several studies on the stellar mass function \citep[see, e.g.,][]{Santini_2022, Navarro-Carrera_2024}, the Initial Mass Function \citep[see, e.g.,][]{Dabringhausen, Kirkpatrick_2024, Tanvir}, the stellar-to-halo mass ratio \citep[see, e.g.,][]{Mitchell, Shuntov, Danieli_2023}, the stellar mass-metallicity relationship  \citep[see, e.g.,][]{Leethochawalit_2019, Baker, Chartab_2024}, or the relationship between the stellar mass and the star formation rate \citep[see, e.g.,][]{Matthee, Pearson, Cooke_2023}. Stellar mass is also used in studying GWs detected with Pulsar Timing Arrays (PTAs), which operate in the nanohertz frequency regime \citep[see, e.g.,][]{PTA_paper}. Researchers approximate the mass of a galaxy's bulge using the $M_\ast$ estimated by the method of \citet{Cappellari_2013} and then utilize the bulge-mass to supermassive black hole-mass relation to calculate the mass of the central black hole \citep{PTA}.

Stellar mass estimation is a complex task based on a galaxy's emitted light \citep[see][for a detailed review]{Courteau_2014}.
A galaxy can be modelled as an ensemble of simple stellar populations \citep[e.g.,][]{SSP} comprising stars of the same age, similar composition, and kinematics, such as globular or open clusters. An essential property of simple stellar populations is the Initial Mass Function \citep[IMF, e.g.,][]{Salpeter, Kroupa, Chabrier_2003}, which describes the fraction of stars with different masses at birth and is typically determined through observations. The Star Formation History (SFH) represents the distribution of stars over time and chemical enrichment \citep{Leja, Lower_2020}. 
Galaxy spectra can be modelled using evolutionary population synthesis models, such as those from \citet{BC03} and \citet{Maraston_05}. Assuming an IMF and SFH, we can calculate the contribution of individual stars based on these models and sum them to obtain the galaxy spectrum and stellar mass-to-light ratio ($M_\ast/L$). We can then estimate the stellar masses of real galaxies by comparing these model spectra or spectral energy distributions with the observed ones or by using the stellar mass-to-light ratio. The uncertainties in the theories lead to uncertainties in the $M_\ast$ estimates \citep[e.g.,][]{Mitchell_13, Pacifici}. 

The mid-infrared $M_\ast/L$ is less sensitive to variations in stellar populations, especially in galaxies without ongoing star formation \citep{Rock, Wen}. It is also less affected by age, SFH, and metallicity compared to shorter wavelengths \citep{Meidt_2014}. The WISE \citep{Wright_2010} \textit{W1} band, which predominantly captures light from old stars, is a good tracer of stellar mass, while the \textit{W1--W2} colour is independent of the age and mass function of the stellar population \citep{Jarrett_2013}. According to \citet{Kettlety_18}, $M_\ast$ estimates based on \textit{W1} luminosity can be as accurate as those from SED fitting methods for passive galaxies.
Several studies \citep*[see, e.g.,][]{Brinchmann_2000, Cole_2001, Bundy_2005} found that near-infrared emissions \citep[like the $K_s$ band of 2MASS, see][]{Skrutskie_2006} are ideal for SME because they are less absorbed by the dust and less dependent on SFH. However, modelling the light from the thermally pulsing asymptotic giant branch is uncertain, and some models do not include it \citep[see, e.g.,][]{BC03}. This phase can significantly impact near- and mid-infrared luminosities \citep*{Conroy_2009, Meidt_2012}, with SME uncertainties reaching up to 0.6 dex when these stars dominate the spectrum \citep{Conroy_2009}.

In this paper, we test methods based on infrared (IR) luminosities and their applicability in GW astronomy. Methods from \citet{Jarrett_2013} and \citet{Cluver_2014} are based on the \textit{W1-W2} colour and the \textit{W1} luminosity; the method from \citet{Kettlety_18} utilizes only \textit{W1} luminosity; and \citet{Cappellari_2013} relies on $K_s$-band magnitude.
First, we implement these SME methods and compare them to the robust $M_\ast$ estimates from Galaxy and Mass Assembly \citep[GAMA; see][]{GAMA} Data Release 3 \citep{Baldry_2018}, which are derived using SED fitting, to assess their accuracy. Then, we rank potential host galaxies of CBCs by combining their localisation probability and their $M_\ast$ from the tested methods to evaluate the effectiveness of these methods in GW astronomy.

The paper is organized as follows. Section$~$\ref{sec:Data} describes the galaxy catalogues and data used in this study. Section$~$\ref{sec:Methods} introduces our analysis methods, with Section$~$\ref{sec:IRmethods} detailing the four luminosity-based SME methods, Section$~$\ref{sec:method:compare} summarising the methods we used to compare the resulting $M_\ast$ values and those from the GAMA catalogue, and Section$~$\ref{sec:method:rank} presenting the methods for ranking possible host galaxies of CBC events based on their stellar masses and describing the generation of simulated CBC events. Section$~$\ref{sec:results} presents our results on the accuracy and applicability of these methods, with Section$~$\ref{sec:compare} comparing the results of these tested methods to the reference values from GAMA and Section$~$\ref{sec:weight} showing our rankings based on different SME methods. We conclude in Section$~$\ref{sec:Conclusions}.

\section{Data}\label{sec:Data}
 This section describes the data used for the tested IR luminosity-based SME methods (see Section$~$\ref{sec:IRmethods}). Photometric data originates from the 2MASS and WISE surveys, redshift data is taken from the GLADE+ catalogue. 
 We also introduce the GAMA survey and its stellar mass estimates, used as references in Section$~$\ref{sec:compare}, and describe the sample of galaxies used in our analysis.

\subsection{Data for stellar mass estimation}
The Two Micron All Sky Survey (2MASS) observed nearly the entire sky in the mid-infrared \textit{J} (1.25$~\mu$m), \textit{H} (1.65$~\mu$m) and $K_s$ (2.16$~\mu$m) bands \citep{Skrutskie_2006}. The 2MASS Extended Source Catalog (XSC) contains 1.6 million extended objects, about 97 per cent of which are galaxies, with the remaining being nebulae in the Milky Way. The algorithms used to create the XSC are described in \citet{Jarrett_2000}. Our study utilizes the \textit{k\_m\_ext} magnitude, which is the total $K_s$ band magnitude obtained by using the 'total' aperture. This aperture is derived by extrapolating the radial surface brightness from the standard isophote to a scale length that covers the deduced extent of the galaxy \citep[for details, see section$~$IV.5.e. of][]{2MASS_exp}.

The Wide-field Infrared Survey Explorer (\textit{WISE}) is a space telescope that mapped the entire sky in four mid-infrared bands at 3.4$~\mu$m (\textit{W1}$~$band),  4.6$~\mu$m (\textit{W2}$~$band), 12$~\mu$m (\textit{W3}$~$band) and 22$~\mu$m (\textit{W4}$~$band) wavelengths \citep{Wright_2010}. The \textit{W1} band is most sensitive to stellar light, while the \textit{W2} band is also sensitive to hot dust, using the (\textit{W1-W2}) colour, we can distinguish active galactic nuclei from galaxies  \citep{Jarrett_2011, Stern_2012}. 
The Image Atlas \citep{Atlas_exp} provides access to the  WISE frames, while the AllWISE data release \citep{AllWISE_exp} offers the photometry for 747 million sources, mostly stars and galaxies. 
In WISE photometry, the background (like distant galaxies and scattered light)  increases noise, making \textit{point spread function} (PSF) photometry recommended \citep{marsh_jarrett_2012}. PSF photometry is optimal for unresolved (point-like) sources, but resolved sources like nearby galaxies are measured as the sum of several point sources, or their fluxes are highly underestimated \citep{Cluver_2014}. The AllWISE data release provides \textsl{w*mpro} magnitudes, derived using PSF photometry, with * indicating the number of the band. 

The WISE$\times$SuperCOSMOS catalogue \citep{Bilicki_2016} was created by cross-matching the AllWISE and the optical SuperCOSMOS catalogue \citep{Hambly}. This sample contains more than 20 million galaxies and is approximately 95 per cent pure. \cite{Bilicki_2016} addressed the above-mentioned issues of the  WISE magnitudes by filtering out the bright ($W1 <$ 13.8)  sources,  ensuring that only galaxies below the resolution threshold remained. This approach provides well-determined magnitudes by PSF photometry, and they also corrected the magnitudes for the galactic extinction according to \citet{Schlafly_2011}.

Our goal is to estimate stellar masses using stellar mass-to-light ratios, which requires the luminosities of galaxies, and therefore we need to have luminosity distances and redshifts aside from apparent magnitudes. However, neither the 2MASS XSC nor the AllWISE catalogue includes such data. \citet{Bilicki_2013} utilized the neural network-based ANNz algorithm to obtain redshifts released in the 2MASS Photometric Redshift (2MPZ) catalogue, and \citet{Bilicki_2016} applied the same method to the  WISE$\times$SuperCOSMOS catalogue. We used the luminosity distance data from the GLADE+ catalogue\footnote{GAMA also contains redshift data. However, we aim to compare the estimates based on limited data with more sophisticated values. The GLADE+ includes a mix of spectroscopic and photometric redshifts, representing a more practical case.} \citep{Dalya_22}, which incorporates the 2MASS and WISE$\times$SuperCOSMOS catalogues.  GLADE+ galaxies with 2MASS $K_s$ and WISE magnitudes allow for comparison of SME methods based on data from different surveys. The redshift data in the GLADE+ catalogue may also originate from the HyperLEDA \citep{Makarov_14} and the GWGC \citep{White_2011}. GLADE+ uses a flat $\Lambda$CDM cosmology with parameters from \citet{Planck18}: $H_0 = 100 h = 67.66 ~ \text{km} ~ \text{s}^{-1} ~ \text{Mpc}^{-1}$, $\Omega_M $ = 0.3111, and $\Omega_\Lambda =  0.6889.$

\subsection{Reference stellar masses}
We compared the results of different SME methods to the stellar masses of the Galaxy And Mass Assembly (GAMA) Data Release\footnote{GAMA has a newer data release (DR4), which is updated with new versions over time (\url{https://www.gama-survey.org/dr4/}). For reproducibility, we used DR3.}$~$3 \citep[DR3,][]{Baldry_2018}.   The GAMA survey\footnote{\url{http://www.gama-survey.org/}} \citep{GAMA} aimed to study cosmology and galaxy formation and evolution using a multi-wavelength photometric and spectroscopic dataset. 
\citet{Cluver_2014} matched the AllWISE counterparts of the GAMA galaxies in the G12 and G15 regions, and later it was done for the G09 region too. Thus, we identified GAMA counterparts for our galaxies using their \textit{wiseX} IDs.

The GAMA DR3 provides three different SMEs for the galaxies.
The StellarMasses and the StellarMassesLambdar tables contain $M_\ast$ values estimated by \citet{Taylor_11}, using the evolutionary models of \citet{BC03} with the IMF of \citet{Chabrier_2003}, an exponentially declining SFH, and the dust extinction law of \citet{Calzetti_2000}. 
The difference between these tables is that the StellarMasses\footnote{StellarMassesv19, \url{https://www.gama-survey.org/dr3/schema/table.php?id=43}} used aperture-matched photometry derived with Source Extractor \citep{SExtractor}, while the StellarMassesLambdar\footnote{StellarMassesv20, \url{https://www.gama-survey.org/dr3/schema/table.php?id=44}} table is based on the photometry derived using the Lambda Adaptive Multi-Band Deblending Algorithm in R (\textsc{lambdar}) program \citep{LAMBDAR}. 
Both tables contain the base-10 logarithm of the stellar mass (\textsl{logmstar})  and we applied aperture correction to obtain the values used in our analysis.\footnote{Aperture correction is described here: \url{https://www.gama-survey.org/dr3/schema/dmu.php?id=9}. We retained galaxies with $1/3 < \text{fluxscale} < 3$.}
Unrealistically low ($M_\ast < 10^6 ~\text{M}_\odot$) and high ($M_\ast > 10^{13} ~\text{M}_\odot$) values were excluded from our analysis.\footnote{See \url{https://www.gama-survey.org/dr3/schema/dmu.php?id=9}.} 

The $M_\ast$ values in the MagPhys table\footnote{\url{http://www.gama-survey.org/dr3/schema/table.php?id=82}} are also based on \textsc{lambdar} photometry. Galaxy parameters were obtained by running the \textsc{magphys} SED-fitting code \citep{MAGPHYS}, which utilizes an updated version of the \citet{BC03} stellar population synthesis code that accounts for the effect of the thermally pulsing asymptotic giant branch \citep{bruzual_2006}. It assumes the IMF of \citet{Chabrier_2003} and an exponentially declining SFH. They used the dust attenuation model of \citet{Charlot_2000}. The \textsl{mass\_stellar\_best\_fit} column gives the stellar mass values in the MagPhys table. We applied the same aperture corrections here as for the StellarMasses and StellarMassesLambdar tables.

We only consider galaxies that have both 2MASS $K_s$,  WISE \textit{W1}, \textit{W2}, redshift data, and $M_\ast$ values from the three GAMA tables. We estimated the stellar masses of these galaxies using the four methods detailed in Section$~$\ref{sec:IRmethods}. A total of 4,291 galaxies met the criteria, with $M_\ast$ values between $10^6$ and $10^{13} ~\text{M}_\odot$. Hereafter, we refer to this set of galaxies as our sample.

\section{Methods}\label{sec:Methods}

In this section, we describe the methods used in our analysis. 
Section$~$\ref{sec:IRmethods} summarizes the origin, implementation, and uncertainty estimation of the four tested SME methods based on IR magnitudes: Jarrett's  \citep{Jarrett_2013}, Cluver's  \citep{Cluver_2014}, Kettlety's \citep{Kettlety_18}, and Cappellari's \citep{Cappellari_2013}. Section$~$\ref{sec:method:compare} describes our approach for comparing the $M_\ast$ estimates from these SME methods with the reference GAMA values.  Section$~$\ref{sec:method:rank} presents the ranking procedures for CBC host galaxies based on their stellar masses, and the simulation of GW observations used to validate these rankings.

\subsection{Stellar mass estimation methods based on IR magnitudes}\label{sec:IRmethods}
\subsubsection{Jarrett's method}
\cite{Jarrett_2013} studied 17 nearby galaxies with diverse morphologies. They reconstructed the WISE Atlas images with  MCM-HiRes to improve the resolution by a factor of three or four \citep{Jarrett_2012}, then measured the fluxes of the galaxies.
Using equation$~$(5) from \cite{Zhu_2010}, the ratio of stellar mass and 2MASS $K_s$ band luminosity can be estimated from the SDSS $g-r$ colour as: 
\begin{equation}\label{eq:Zhu}
    \log_{10} \frac{M_\ast}{\nu L_\nu(K_s)} = -1.29 (\pm 0.05) +1.42(\pm 0.06) (g-r).
\end{equation}
\cite{Jarrett_2013} used this SME to calculate the $M_\ast/L_{W1}$ ratio in $\text{M}_\odot/\text{L}_\odot$ units (where $\text{L}_\odot$ is the total solar luminosity, $3.839 ~ \times ~ 10^{33} ~\text{erg}~\text{s}^{-1}$, and $\text{M}_\odot$ is the mass of the Sun). They observed a linear trend between this stellar mass-to-light ratio and the \textit{W1-W2} colour \citep[see equation$~$9 in][]{Jarrett_2013}: 
\begin{equation}\label{eq:Jarrett_original}
    \log_{10} (M_\ast(K_s)/L_{W1}) = -0.25(\pm0.03)-2.1(\pm0.2)(W1-W2).
\end{equation}
$M_\ast(K_s)$ marks that the stellar mass is estimated using the 2MASS $K_s$ in-band luminosity and the relation in equation$~$(\ref{eq:Zhu}).
The \textit{W1} in-band luminosity \citep[measured relative to the Sun in the given band, see equation$~$3 in][]{Jarrett_2013} is given by
\begin{equation}\label{eq:L_W1}
    L_{W1} = 10^{-0.4(M_{W1}-M_{\odot, W1})},
\end{equation}
where $M_{W1}$ is the absolute \textit{W1} band magnitude of the galaxy and $M_{\odot, W1} = 3.24$ mag is the absolute in-band magnitude of the Sun from \citet{Jarrett_2013}.
We calculated the absolute \textit{W1} and \textit{W2} band magnitudes using
\begin{equation}\label{eq:abs_mag}
    M_{\{W1,W2\}} = m_{\{W1,W2\}} + 5 -5 \log_{10} d_L - K_{\{W1,W2\}},
\end{equation}
where $m_{\{W1,W2\}}$ is the apparent magnitude, $d_L$ is the luminosity distance, $K_{\{W1,W2\}}$ is the \textit{k}-correction given by \cite{patrick_2015} as
    $K_{W1} = - 2.5854 ~z$ and $ K_{W2} = - 2.9358 ~ z$. 

\subsubsection{Cluver's method}
\cite{Cluver_2014} crossmatched the galaxies in the G12 and G15 regions of GAMA DR2 with their WISE counterparts. The PSF photometry of the WISE catalogue,  effective for point sources, is unsuitable for resolved sources, as discussed in Section$~$\ref{sec:Data}. To address this, \cite{Cluver_2014} produced new images using the Variable-Pixel Linear Reconstruction algorithm to obtain the isophotal photometry for resolved sources and provided guidelines for selecting the best photometry in cases of low signal-to-noise ratio. They determined rest-frame colours by building a SED from optical, near-infrared, and mid-infrared flux densities and fitting an empirical template library from \cite{Brown_2014}.
The GAMA DR2 contains the stellar mass estimates (StellarMassesv15) from \cite{Taylor_11} based on \citet{BC03} models. These estimates are the most accurate for $z<0.15$, which is why \cite{Cluver_2014} studied galaxies within this redshift range. 
They used galaxies with high signal-to-noise ratios (S/N > 13.5 in \textit{W1} and \textit{W2}) to fit the observed linear relationship between the stellar mass-to-light ratio and the rest frame \textit{W1-W2} colour with the maximum likelihood method. Their fit shows that passive galaxies have larger $M_\ast/L$ than star-forming galaxies. For the entire sample, they obtained the following relation:
\begin{equation}\label{eq:Cluver_original}
    \log_{10} (M_\ast / L_{W1}) = -1.96 (W1-W2)-0.03,
\end{equation}
where $L_{W1}$ is the \textit{W1} band luminosity (see equation$~$\ref{eq:L_W1}), \textit{W1-W2} is the rest frame colour, which we calculated in our analysis as the difference between \textit{W1} and \textit{W2} absolute magnitudes (equation$~$\ref{eq:abs_mag}). 
They note that their results are highly consistent with the \citet{Jarrett_2013}.
Fig.$~$8 in \citet{Cluver_2014} illustrates the fitting accuracy, but since parameter errors were not provided, we cannot include this in our error estimation. 

\subsubsection{Kettlety's method}
\citet{Kettlety_18} compared stellar masses estimated from mid-infrared bands with those from an optical--to--near-infrared SED fitting. They used photometry from GAMA DR2 \citep{Liske}, with WISE data added by \citet{Cluver_2014}.  The StellarMassesv18 values in GAMA DR2 were determined by SED fitting using SDSS \textit{ugriz} and VISTA-VIKING \textit{ZYJHK} bands, based on \citet{BC03} simple stellar population evolutionary models and \citet{Chabrier_2003} IMF. Composite stellar populations were obtained by weighting these simple stellar populations according to a SFH that declines with a time constant $\tau$ and begins at $t$ time before the observation time. Dust extinction was assumed to be uniform and described by the \citet{Calzetti_2000} extinction law. This method is an updated version of the SED fitting of \citet{Taylor_11}. 
\citet{Kettlety_18} studied galaxies with spectroscopic redshift $z \leq 0.15$ to remove and minimize  uncertain mid-IR \textit{k}-corrections, while the $z>0.003$ condition excluded possible stellar contamination. They selected passive galaxies and galaxies with old stellar populations by cleaning their sample based on markers of recent star formation in the GAMA catalogue and WISE colours.
\citet{Kettlety_18} determined \textit{k}-correction as:
\begin{equation}\label{eq:W1_Kcor}
    K = -7.1 \log_{10} (1+z).
\end{equation} 
They found that for low-redshift, passive galaxies, the stellar mass-luminosity ratio is
\begin{equation}\label{eq:Kettlety:passive}
    M_\ast/L_{W1} = 0.65 \pm 0.07,
\end{equation}
where the error does not contain the uncertainty of the population synthesis models. This ratio increases with age, but determining the age requires SED fitting, which is not possible with only a few magnitude bands.
When including star-forming galaxies in their sample, \citet{Kettlety_18} found a lower  stellar mass-to-light ratio:
\begin{equation}\label{eq:Kettlety:active}
    \log_{10} (M_\ast/L_{W1}) \simeq -0.4 \pm 0.2,
\end{equation}
corresponding to $M_\ast/L_{W1} \simeq 0.398$. 
We classified galaxies as active if $W2 - W3 \leq 1.5$ and passive otherwise. The \textit{W2}$~$band is sensitive to evolved stellar populations and hot dust, while the \textit{W3}$~$band, dominated by the 11.3 $\mu$m polycyclic aromatic hydrocarbon feature and dust continuum, and sensitive to star formation. Thus, the \textit{W2 - W3} colour can distinguish between old stellar population-dominated galaxies and star-forming galaxies \citep{Cluver_2014, Jarrett_2011}.
For active galaxies and those without \textit{W3} data, we used equation$~$(\ref{eq:Kettlety:active}) to calculate $M_\ast$, and for passive galaxies, we used equation$~$(\ref{eq:Kettlety:passive}).
To apply equations$~$(\ref{eq:Kettlety:passive})$~$and$~$(\ref{eq:Kettlety:active}), we calculated the \textit{W1} band luminosity (equation$~$\ref{eq:L_W1}) using the magnitude definition in equation$~$(\ref{eq:abs_mag}) and the \textit{k}-correction approximation in equation$~$(\ref{eq:W1_Kcor}). 
\citet{Kettlety_18} found that for passive galaxies, SED fitting may not be necessary for SME, as good approximations can also be obtained from \textit{W1} luminosities.

\subsubsection{Cappellari's method}
The previous techniques rely on evolutionary population synthesis models. In contrast, the SME method introduced in \citet{Cappellari_13} uses a dynamical modeling approach.
They employed the \textit{Jeans Anisotropic Multi-Gaussian Expansion} (JAM) models, which require the surface brightness distribution and the first two velocity moments as input. JAM models accurately describe the galaxy photometry and reproduce the observed kinematics of galaxies. One of the key output parameters is the \textit{total} mass-to-light ratio within a sphere with radius equal to the galaxy's \textit{projected} half-light radius ($R_e \equiv \sqrt{A_e/\pi}$, where $A_e$ is the area of the effective isophote containing half of the total light), denoted as $(M/L)_{\text{JAM}}$. 
As explained in section$~$4.3 of \citet{Cappellari_13}:
\begin{equation}
    M_\text{JAM} \equiv L  (M/L)_\text{JAM} \approx 2  M_{1/2} \approx M_\ast,
\end{equation}
where $M_\text{JAM}$ is the dynamical mass of the galaxy (including both bright and dark matter) within a sphere of radius $R_e$,  $L$ is the total luminosity of the galaxy, $M_{1/2}$ is the total mass inside the sphere of radius $r_{1/2}$, enclosing half of the total light ($r_{1/2} \equiv \sqrt[3]{3V_e/(4\pi)}$, where $V_e$ is the volume of the isosurface containing half of the total light, 
$r_{1/2} \approx 1.33 R_e$ for spherical models). The approximation $2 M_{1/2} \approx M_\ast$ holds because the contribution of dark matter within the $r_{1/2}$ sphere is relatively small, making this approximation generally accurate within 20 per cent.
Using 743 galaxies from the ATLAS$^\text{3D}$ sample \citep{ATLAS3D}, which are closer than $d_L = 42~$Mpc and their $K_s$ band magnitude is $M_K<-21.5~$mag ($M_\ast \geq 6 \times 10^9 ~\text{M}_\odot$), \citet{Cappellari_2013} found a linear relationship between $\log_{10} M_\ast$ and $M_K$:
\begin{equation}\label{eq:Cappellari}
    \log_{10} M_\ast \approx (10.583 \pm 0.009) - (0.445 \pm 0.009)(M_K+23).
\end{equation}
When applying this method we calculated the $M_K$ according to \citet{Ma_2014}:
\begin{equation}\label{eq:MKs}
    M_K = m - 5 \log_{10} d_L - 25 -0.11  A_V,
\end{equation}
where  $m$ represents the $K_s$ apparent magnitude from fit extrapolation (\textit{k\_m\_ext}) of the 2MASS catalogue, $d_L$ is the luminosity distance measured in megaparsecs, and $A_V$ denotes the Landolt V galactic extinction:
\begin{equation}\label{eq:AV}
    A_V = 2.742 ~ E(B-V)_\text{SFD},
\end{equation}
where $E(B-V)_\text{SFD}$ is the reddening computed by the \textsc{dustmaps} package \citep{Green2018} using the SFD map of \citet*{Schlegel_1998}, and the coefficient is necessary for the conversion of the $E(B-V)_\text{SFD}$ values to the magnitudes of $A_V$. This conversion coefficient was obtained from table$~$6 of \citet{Schlafly_2011} in the case of $R_V = A_V / E(B-V) = 3.1$.
For \textit{k}-correction, we used $ K_K=~-1.8178 ~z$ from \citet{patrick_2015}.
It is important to note that this method of SME is fitted for early-type galaxies, and it may not be accurate for spirals. \citet{Cappellari_12} cautions that commonly used IMFs should not be applied universally across all galaxy types.  In early-type galaxies, stellar masses from evolutionary population synthesis models with improperly chosen IMFs may differ by up to three orders of magnitude from dynamical model results.

\subsubsection{Error estimation}
We used error propagation to calculate the uncertainties of the $M_\ast$ values, which were estimated using the four tested methods.
For magnitudes without provided uncertainties, we used the mean plus two sigma of relative errors: 0.005 for \textit{W1}, 0.01 for \textit{W2}, and 0.02 for $K_s$. The missing uncertainties of luminosity distances were estimated from redshift errors.
From Table$~$\ref{tab:err}, which summarizes the error estimates, we can see that Cappellari's method has the smallest uncertainty for the whole GLADE+ sample. Kettlety's method shows lower errors for passive galaxies than active ones, because the \textit{W1} magnitude mainly traces the older stellar population that is dominant in passive galaxies, and does not provide information about the younger stars present in active galaxies. For Cluver's method, the parameter errors in equation$~$(\ref{eq:Cluver_original}) is not provided, thus, they are not considered.

\begin{table*}
\caption{Estimated relative errors of the $M_\ast$ values in per cents from different SME methods (see Section$~$\ref{sec:IRmethods}), with mean values followed by medians in brackets. Statistics are provided for the entire GLADE+ sample and our sample of 4,291 galaxies (see Section$~$\ref{sec:Data}). For Kettlety's method, errors are reported separately for passive and active galaxies. We arranged the columns in ascending order based on the errors obtained in the GLADE+ sample.}\label{tab:err}
\centering
\begin{tabular}{|c|c|c|c|c|c|}
\hline
                    & \textbf{Cappellari} & \textbf{\begin{tabular}[c]{@{}c@{}}Kettlety\\ passive galaxies\end{tabular}} & \textbf{Cluver} & \textbf{Jarrett} & \textbf{\begin{tabular}[c]{@{}c@{}}Kettlety\\ active galaxies\end{tabular}} \\ \hline
\textbf{GLADE+}     & 27 (24)             & 32 (37)                                                                      & 51 (44)         & 52 (48)          & 58 (63)                                                                     \\ 
\textbf{Our sample} & 20 (18)             & 15 (12)                                                                      & 15 (7)          & 20 (13)          & 49 (46)                                                                     \\ \hline
\end{tabular}
\end{table*}

We calculated the error ($\epsilon$) for stellar mass estimates from GAMA DR3 using 
\begin{equation}
    \epsilon= \frac{1}{N} ~ \sum\limits_i \frac{\sigma_i}{\mu_i},
\end{equation}
where $i$ denotes the $i$th galaxy in our sample of 4,291 galaxies (see Section$~$\ref{sec:Data}), $\sigma_i$ is the standard deviation of the three $M_\ast$ values from the StellarMasses, the StellarMassesLambdar, and the MagPhys tables, $\mu_i$ is the mean of them, and $N$ is the number of galaxies. We found that the value of $\epsilon$ is $\sim$10 per cent ($\sim$0.04 dex).  
The average deviation from the mean stellar mass is 0.04 dex ($\sim$10 per cent), with a maximum deviation of 0.36 dex ($\sim$130 per cent). However, $M_\ast$ based on different SED fitting methods can differ by up to 0.6 dex ($\sim$ 300 per cent) due to varying assumptions in SPS models \citep[see, e.g.,][]{Conroy, Courteau_2014}.

\subsection{Comparison with GAMA stellar masses}\label{sec:method:compare}
We examined the relationship between stellar masses from the four tested methods (see Section$~$\ref{sec:IRmethods}) and those from the GAMA catalogue. Assuming linear relationships, we calculated the Pearson correlation coefficients, accounting for uncertainty in $M_\ast$ values using Monte Carlo simulations. 
We compared the $\log_{10} M_\ast$ values from each tested method with the GAMA reference values and fitted the observed relationships using the least squares method.

To assess the accuracy of the luminosity-based SME methods, we defined the  root mean squared error as
\begin{equation}\label{eq:chi2}
D = \sqrt{\frac{1}{N} \sum\limits_{i} (x_i-m_i)^2},
\end{equation}
where $x_i$ is the $\log_{10}M_\ast$ of the $i$th galaxy in our sample calculated with the tested method, $m_i$ is the corresponding $\log_{10}M_\ast$ from the GAMA catalogue (assumed to be the true value), and $N$ is the total number of galaxies. A smaller $D$ indicates better agreement between the values from the tested methods and those from the GAMA catalogue. The results of the comparisons are detailed in Section$~$\ref{sec:results}.

\subsection{Ranking host galaxies according to stellar masses}\label{sec:method:rank}
Ranking potential host galaxies of CBC events based on their $M_\ast$ is an important application of stellar mass in GW astronomy. We used the ranking to evaluate the effectiveness of the tested luminosity-based SME methods.
We ranked the potential hosts  based on their localisation probability as detailed in \citet{Singer_2016}. We employed two approaches to incorporate stellar mass into the ranking process. The first, from \citet{Ducoin}, calculates the probability of a galaxy being the CBC host using 
\begin{equation}\label{eq:weight}
G_\text{tot} = P_\text{pos} \times G_\text{mass}, 
\end{equation}
where $P_\text{pos}$ is the probability based on localisation \citep[described in][]{Singer_2016}, and 
\begin{equation}\label{eq:gmass}
G_\text{mass} = \frac{M_{\ast, i}}{\sum\limits_{i}M_{\ast, i}},
\end{equation}
where index $i$ represents the $i$th galaxy within the localisation volume. In a real galaxy catalogue, some galaxies may lack estimated $M_\ast$, making equation$~$\ref{eq:weight} inapplicable. To address this, \citet{Ducoin} redefined $G_\text{tot}$ as
\begin{equation}\label{eq:weight2}
    G_\text{tot} = P_\text{pos}(1+\alpha G_\text{mass}),
\end{equation}
where $\alpha$ ensures equal contribution from both terms:
\begin{equation}\label{eq:alpha}
    \alpha = \frac{\sum P_\text{pos}}{\sum P_\text{pos} G_\text{mass}},
\end{equation}
where the sum is over all galaxies in the localisation volume that have stellar mass. We applied equation$~$\ref{eq:weight2} for ranking the host galaxy of GW170817 \citep{GW170817}.

The second method, from \citet{Artale_2020}, combines the results from the population synthesis code \textsc{Mobse} with galaxy catalogues from the \textsc{EAGLE} cosmological simulation to determine the host galaxy probability for three types of CBCs (binary black hole, binary neutron star, and black hole-neutron star mergers) in four redshift ranges ($z \in [0, 0.1]$, $z \in [0.93, 1.13]$, $z \in [1.87, 2.12]$, and $z \in [5.73, 6.51]$) by using a grid of 25$\times$25$\times$25 points in the stellar mass--SFR--metallicity space. They found that the correlation between host galaxy probability and stellar mass is not strictly linear. We used their table\footnote{\url{https://github.com/mcartale/HostGalaxyProbability/blob/master/BBH/1DHostGalaxyProbability_Mstar_BBH_z0p1.dat}} for binary black holes (BBHs) at $z \in [0, 0.1]$ (the closest redshift range to our maximum redshift of 0.22).
We used linear interpolation to determine the corresponding probability $p(M_\ast)$ (see Appendix$~$\ref{app:Artale_table}) and substituted it for $M_{\ast, i}$ in equation$~$(\ref{eq:weight}).

In Ducoin’s approach, the ranking depends on the relative contribution of each galaxy’s stellar mass, as a result, methods with fixed stellar mass-to-light ratios, such as Kettlety’s and Cappellari’s, can effectively be understood as ranking by W1 or K-band luminosities. However, this does not hold for Artale’s approach, where absolute stellar mass values directly determine the assigned host probabilities.

Alongside GW170817, we tested host galaxy ranking techniques using simulated events. 
We created a mock galaxy catalogue with the Theoretical Astrophysical Observatory\footnote{\url{https://tao.asvo.org.au/tao/}} (TAO) employing the Semi-Analytic Galaxy Evolution (SAGE) model \citep{Croton_2016}. This is a light-cone type catalogue from snapshots of the Millenium N-body simulation \citep{Springel2005} for the whole sky within redshift $z < 0.22$. We calculated the absolute and apparent magnitudes for 2MASS $K_s$, WISE \textit{W1} and \textit{W2} filters using the SED module of TAO with the  \citet{BC03} stellar population model and the \citet{Chabrier_2003} IMF. No intergalactic medium effects or dust attenuation were applied, as they would introduce additional uncertainties.  
We selected galaxies with $M_\ast \geq 10^{8} ~ \text{M}_\odot/h$ to maximize the size of the galaxy catalogue while remaining within the disk capacity limits of TAO, where $h$ is the reduced Hubble constant $h \equiv H_0 \times (100 \text{ km s}^{-1} \text{ Mpc}^{-1})^{-1} = 0.73$. This approach resulted in approximately 78 million galaxies.
We then randomly sampled 1,000 host galaxies weighting them by their stellar masses, assuming that stellar mass correlates with the probability of the galaxy being the host. Additionally, we accounted for the redshift evolution of the merger rate, following the model described by equation$~$(4) of \citet{O3cos}. This model is  based on \citet{MandD} parameterization, which assumes that the binary formation rate follows the SFR. We used $\gamma = 4.56$, $k = 2.86$, and $z_p = 2.47$, as provided in the analysis by \citet{O3cos}.
We simulated one BBH event in each of the 1,000 host galaxies and found that 195 events could be detected with S/N $\geq 11$ by a network of three detectors, two of which are operated with Advanced LIGO O4 sensitivity and one with the Virgo O4 sensitivity\footnote{The noise curves can be downloaded from \url{https://dcc.ligo.org/LIGO-T2000012/public}. We used the aligo\_O4high.txt and avirgo\_O4high\_NEW.txt files.} \citep{Prospects}.
While details such as star formation prescriptions, metallicity dependencies, or environmental effects can influence binary fractions, they do not significantly affect our results. Since we weight host galaxy selection by both stellar mass and the redshift-dependent merger rate evolution (following \citealt{MandD}), both factors contribute to host selection. However, given that our mock galaxy catalogue is limited to 
z<0.22, the redshift evolution of the merger rate does not vary significantly across the sample. In this regime, stellar mass weighting dominates the host galaxy selection, as its variation spans several orders of magnitude. The results of \citet{Artale_1, Artale_2020} demonstrate that at low redshifts, stellar mass is the strongest predictor of BBH merger rate compared to SFR or metallicity.
However, we note that Artale’s ranking approach itself does not assume a strictly monotonic relationship between mass and host probability, allowing us to test how deviations from this assumption impact our results.

We inferred the source properties of the BBH events detected in our simulation  using the Bayesian inference library, \textsc{bilby} \citep{bilby_paper}, commonly used in GW astronomy. These properties include the black hole masses, spins, luminosity distance, right ascension, declination, inclination angle, polarization angle, and phase at coalescence \citep[see table$~$1 of][]{bilby_paper}. We utilized the \textsc{bilby\_pipe} \citep{bilby_pipe_paper} Python tool, which automates the \textsc{bilby} package, along with the \textsc{dynesty} sampler \citep{dynesty, dynesty_code}. The sampler calculates posterior samples and evidence based on input priors and likelihoods. We applied the default priors for BBH mergers \citep[as described in][]{bilby_paper}, which were also used to sample input parameters for the BBH simulation, except for right ascensions, declinations, and luminosity distances, which were fixed to those of the host galaxies. The outputs of the inference include posterior distributions of the source properties and 3D GW localisation probability maps (skymaps) based on the posteriors for right ascension, declination, and luminosity distance.

We cross-matched the resulted skymaps with the simulated galaxy catalogue using the ligo.skymap package\footnote{\url{https://git.ligo.org/lscsoft/ligo.skymap}}. Only events whose true host galaxies lie within the 90 percent localization volume were considered, resulting in a total of 153 events.
We ranked the galaxies based on their localisation probability ($P_\text{pos}$) and then on  $G_\text{tot}$ (see equation$~$\ref{eq:weight}) using different stellar masses: the actual $M_\ast$ values from the simulated galaxy catalogue and estimates from the methods of Jarrett, Cluver, Kettlety, and Cappellari. The comparisons of rankings are presented in Section$~$\ref{sec:weight}.

\section{Results}\label{sec:results}
This section presents our main results. In Section$~$\ref{sec:compare}, we compare the $M_\ast$ values from the tested methods (Section$~$\ref{sec:IRmethods}) with the GAMA stellar masses using the methods detailed in Section$~$\ref{sec:method:compare}. In Section$~$\ref{sec:weight}, we rank potential host galaxies of CBC events using the methods from Section$~$\ref{sec:method:rank} to evaluate the applicability of IR luminosity-based SME methods in GW astronomy.

\subsection{Comparisons with GAMA stellar masses}\label{sec:compare}
\begin{figure}
    \centering
    \includegraphics[scale=0.43]{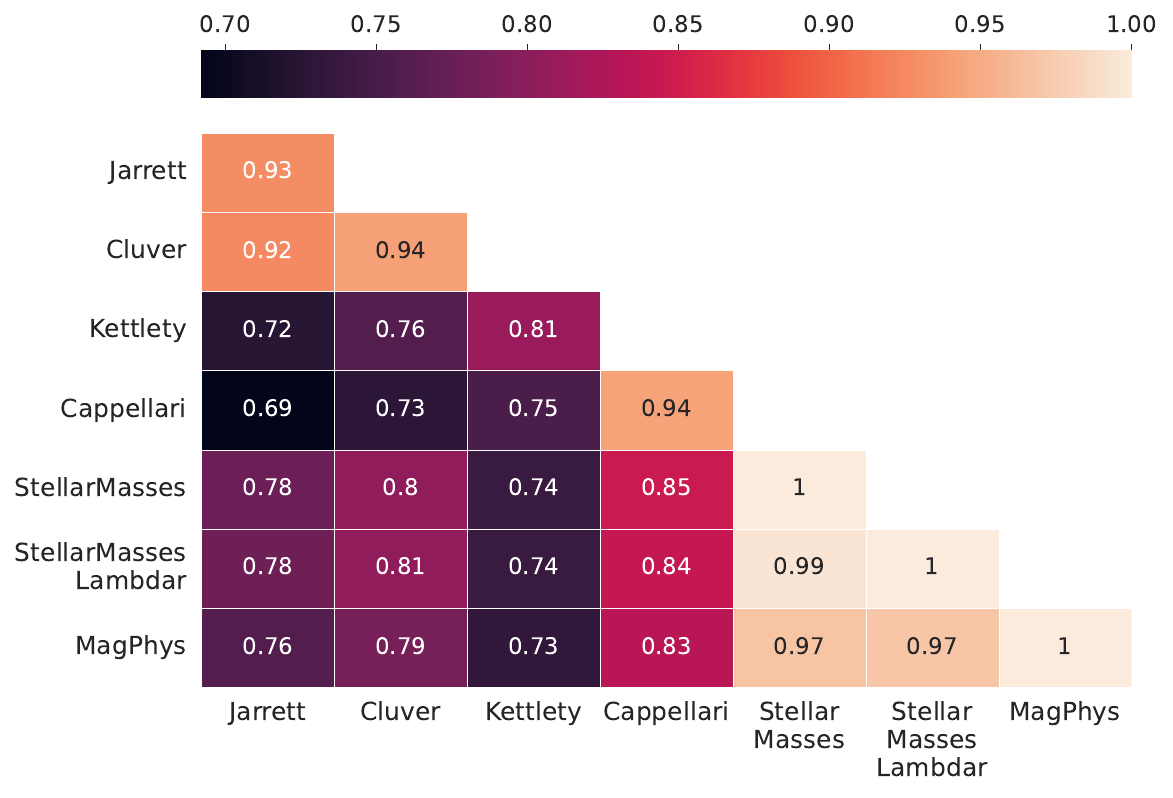} 
    \caption{Pearson correlation coefficients between stellar masses estimated by the four tested SME methods (Jarrett's, Cluver's, Kettlety's, and Cappellari's; see Section$~$\ref{sec:IRmethods} for details) and those from the GAMA catalogue for our sample of 4,291 galaxies. The reference values are from the StellarMasses, the StellarMassesLambdar and the MagPhys tables of the GAMA catalogue. We accounted for uncertainties in $M_\ast$ estimates using Monte Carlo simulations, which is why some coefficients in the diagonal are not equal to 1. Cappellari's method shows the highest correlation with GAMA values, but strong correlations are observed for most methods.} 
    \label{fig:corr}
\end{figure}

We compared the results of the four luminosity-based SME methods to the stellar masses of the GAMA catalogue derived from SED fitting. 
First, we show the correlations of the results from the different SME methods in Fig.$~$\ref{fig:corr}.
We used Monte Carlo simulations to account for uncertainties in the tested methods, sampling from the assumed Gaussian distribution centered on the estimated stellar mass, with the calculated uncertainties of stellar masses (see Section$~$\ref{sec:IRmethods}) assumed to be the standard deviation of the distribution.
This explains why some coefficients in the diagonal are not 1. We observed strong positive correlations (with correlation coefficients greater than 0.7) for nearly all methods, with statistical errors of $\sim5 \cdot 10^{-3}$ from the Monte Carlo simulations. The high correlation coefficient (0.92) between Jarrett's and Cluver's methods is expected, as both use the WISE \textit{W1} luminosity and \textit{W1-W2} colour data  and \citet{Cluver_2014} also demonstrated consistency between them. Cappellari's method shows stronger correlation with the reference GAMA stellar masses compared to the WISE-based methods (Jarrett's, Cluver's, and Kettlety's). 
Furthermore, there is a stronger correlation between Kettlety's and Cappellari's methods than between Kettlety's and the \textit{W1-W2}  methods.
While we observe strong positive correlations between the $M_\ast$ from the tested SME methods and the reference values from the GAMA tables, the correlations between the different GAMA tables themselves (StellarMasses, StellarMassesLambdar, and MagPhys) are even higher, with coefficients of 0.97 and 0.99, indicating very strong agreement. The $M_\ast$ values in the StellarMasses and StellarMassesLambdar tables were obtained using the same SED-fitting method but applied to different photometry, while the $M_\ast$ values in the MagPhys table were derived from the Lambdar photometry (as in the StellarMassesLambdar table), but with a different SED-fitting code (see Section~\ref{sec:IRmethods}).

\begin{figure*}
\subfloat{\includegraphics[width = 3.45in]{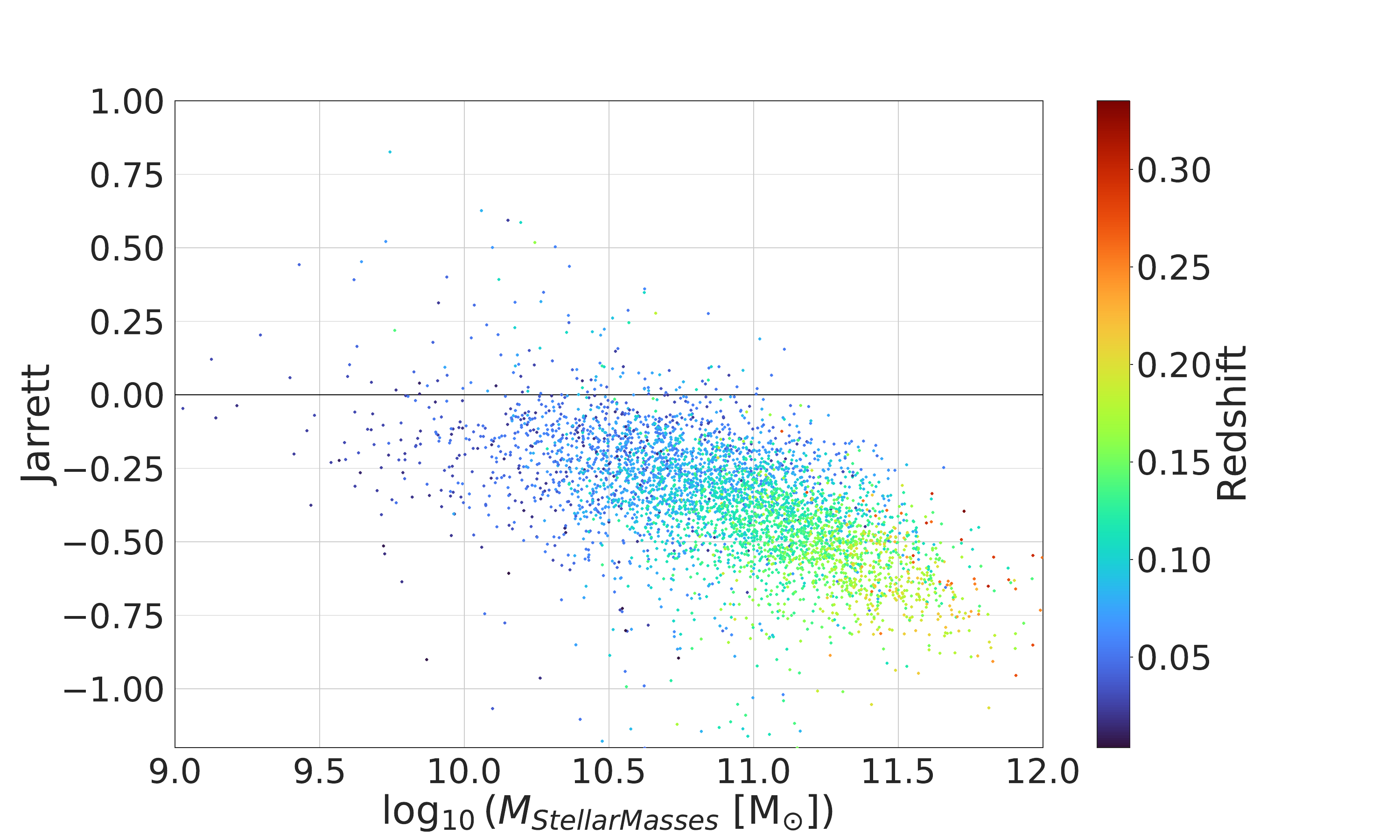}}
\subfloat{\includegraphics[width = 3.45in]{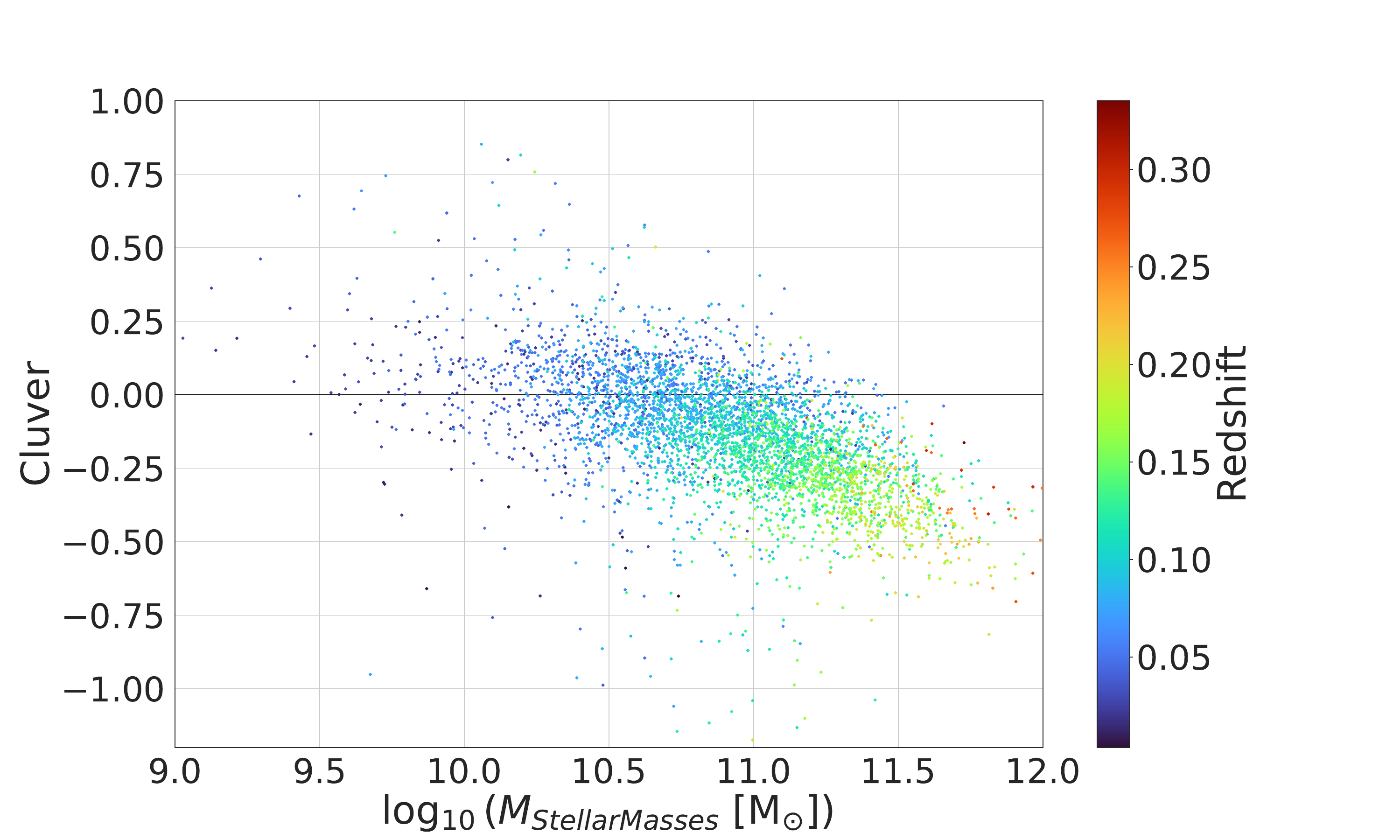}}\\
\subfloat{\includegraphics[width = 3.45in]{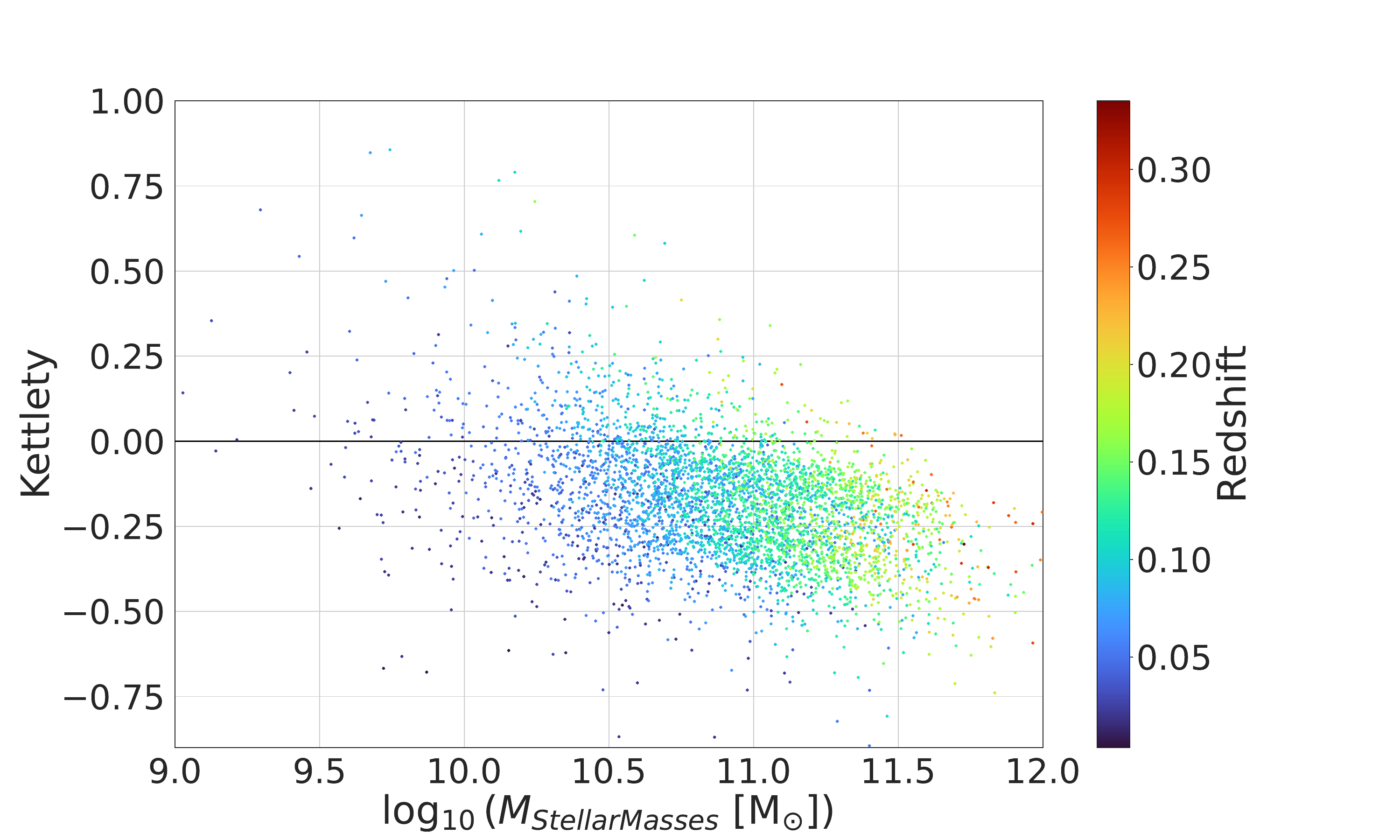}} 
\subfloat{\includegraphics[width = 3.45in]{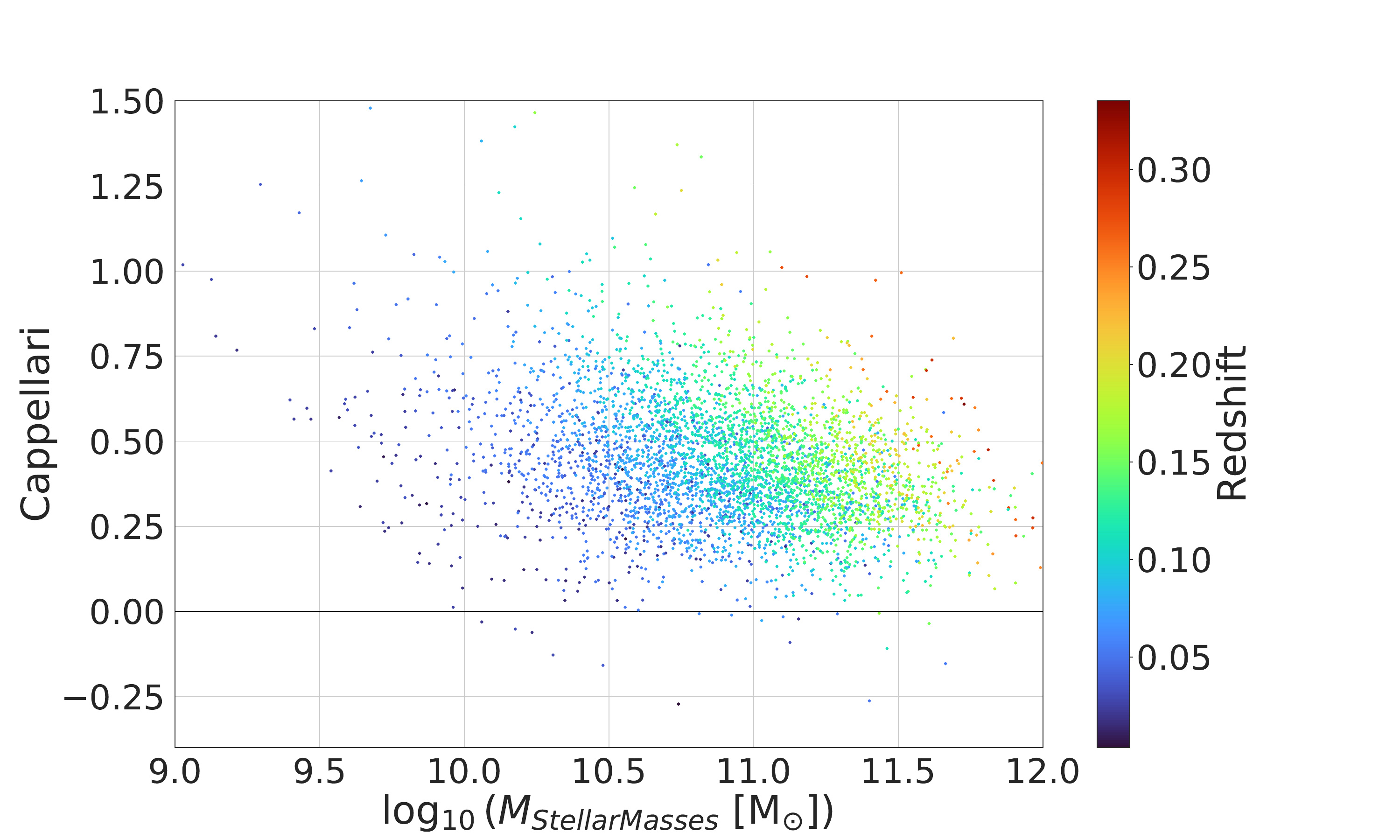}} 
\caption{Logarithmic differences $\log_{10}(M_\text{tested method}/M_\text{StellarMasses})$ between stellar masses derived from the tested methods (see Section$~$\ref{sec:IRmethods}) and the stellar mass estimates in the StellarMasses table of the GAMA catalogue \citep{GAMA}, plotted as a function of the base-10 logarithms of stellar masses from the StellarMasses table. The $y$-axis labels indicate the tested methods, and the colour code represents the redshifts of galaxies in the sample.}
\label{fig:fraction}
\end{figure*}

In Fig.$~$\ref{fig:fraction}, we show the log differences of stellar masses from the tested methods and the StellarMasses table of the GAMA catalogue, plotted against log StellarMasses values.
We obtained similar figures for the other two GAMA tables. The colour scale indicates the redshifts, with more massive galaxies generally at higher redshifts due to magnitude limitations of observations.   For comparison, see Fig.$~$\ref{fig:fraction_GAMA} in Appendix$~$\ref{app:B},  which shows the differences between $M_\ast$ values from different GAMA tables.

We observe that for the WISE-based methods, the differences decrease with increasing StellarMasses values, showing higher $M_\ast$ for lower StellarMasses values, and lower $M_\ast$ for higher StellarMasses values. A similar trend is shown in fig.$~$8 of \citet{Cluver_2014} for galaxies with $z<0.15$. The $z < 0.15$ version of Fig.$~$\ref{fig:fraction} is Fig.$~$\ref{fig:fraction_15} in Appendix$~$\ref{app:diff}. The reason for this declining trend, as also explained in \citet{Cluver_2014}, is that the sample contains more star-forming galaxies  at higher redshifts because they have higher IR luminosities and thus pass the magnitude limitations of the survey more frequently than passive galaxies. However, the WISE 3.4 $\mu$m band is sensitive to light from evolved stars, leading to an underestimation of stellar masses at high redhsifts. Another contributing factor to this trend is the bright-end cut of $W1<13.8$ in \citet{Bilicki_2016}, which excludes low M/L galaxies at high stellar masses and low redshifts (see the sharp diagonal edge in Fig.$~$\ref{fig:fraction}). We note that these selection effects should be considered when interpreting our results.

Fig.$~$\ref{fig:fraction} shows that Cappellari's method generally yields higher stellar masses than the StellarMasses values.
This discrepancy arises because \citet{Cappellari_2013} used early-type (passive) galaxies which have a higher stellar mass-to-light ratio than late-type (active) galaxies \citep{Cluver_2014, Jarrett_2013}. Cappellari's method estimates stellar mass by using the total mass-to-light ratio which may be inadequate. Additionally, \citet{Cappellari_12} suggests that using an IMF, which is determined near the Solar system in evolutionary models, can lead to incorrect SME results. However, these differences do not affect the ranking in Ducoin’s approach, as host galaxy probabilities are derived from relative rather than absolute stellar mass values. In contrast, for Artale’s approach, where absolute mass values determine host probabilities, this systematic offset could introduce biases in ranking results.

\begin{figure*}
\subfloat{\includegraphics[width = 3.45in]{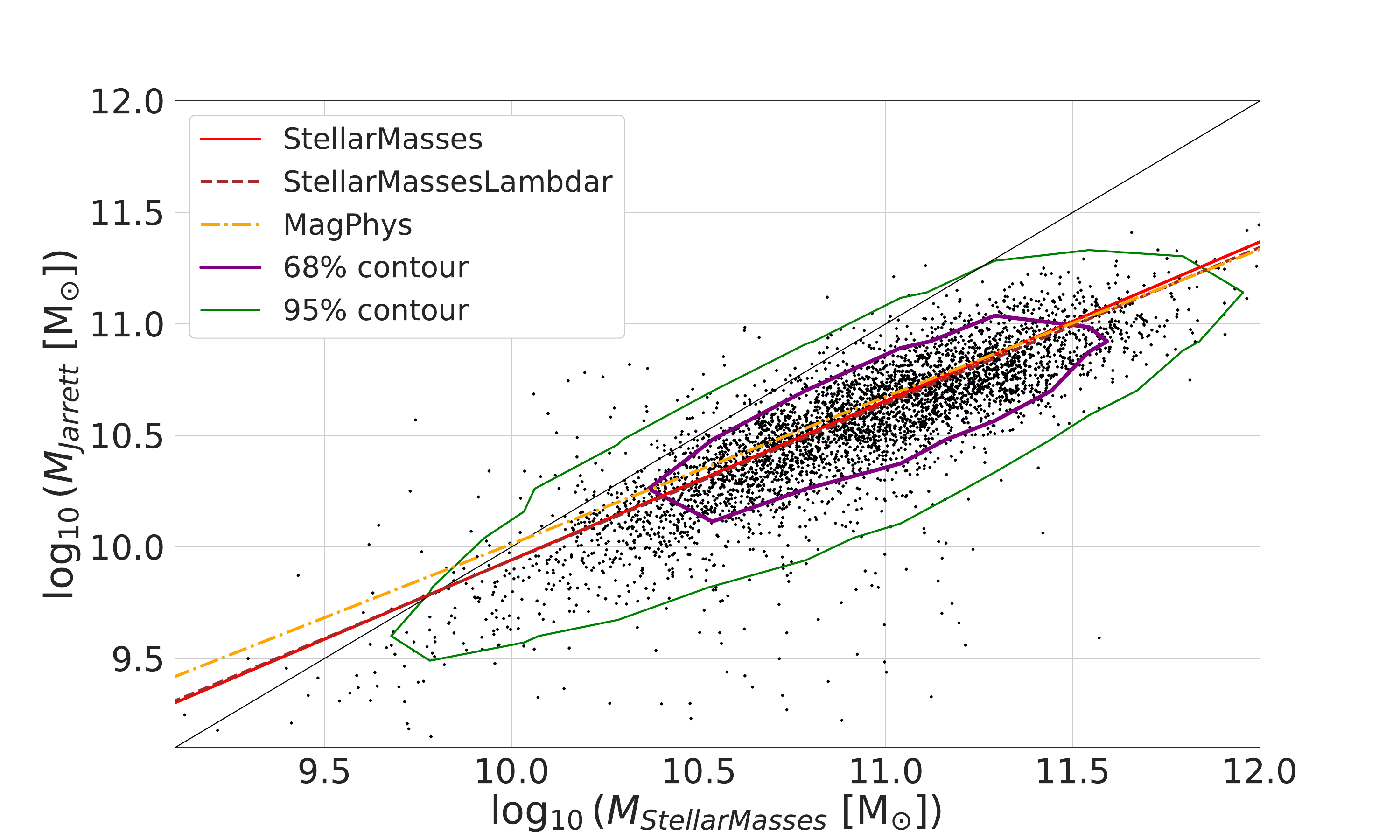}}
\subfloat{\includegraphics[width = 3.45in]{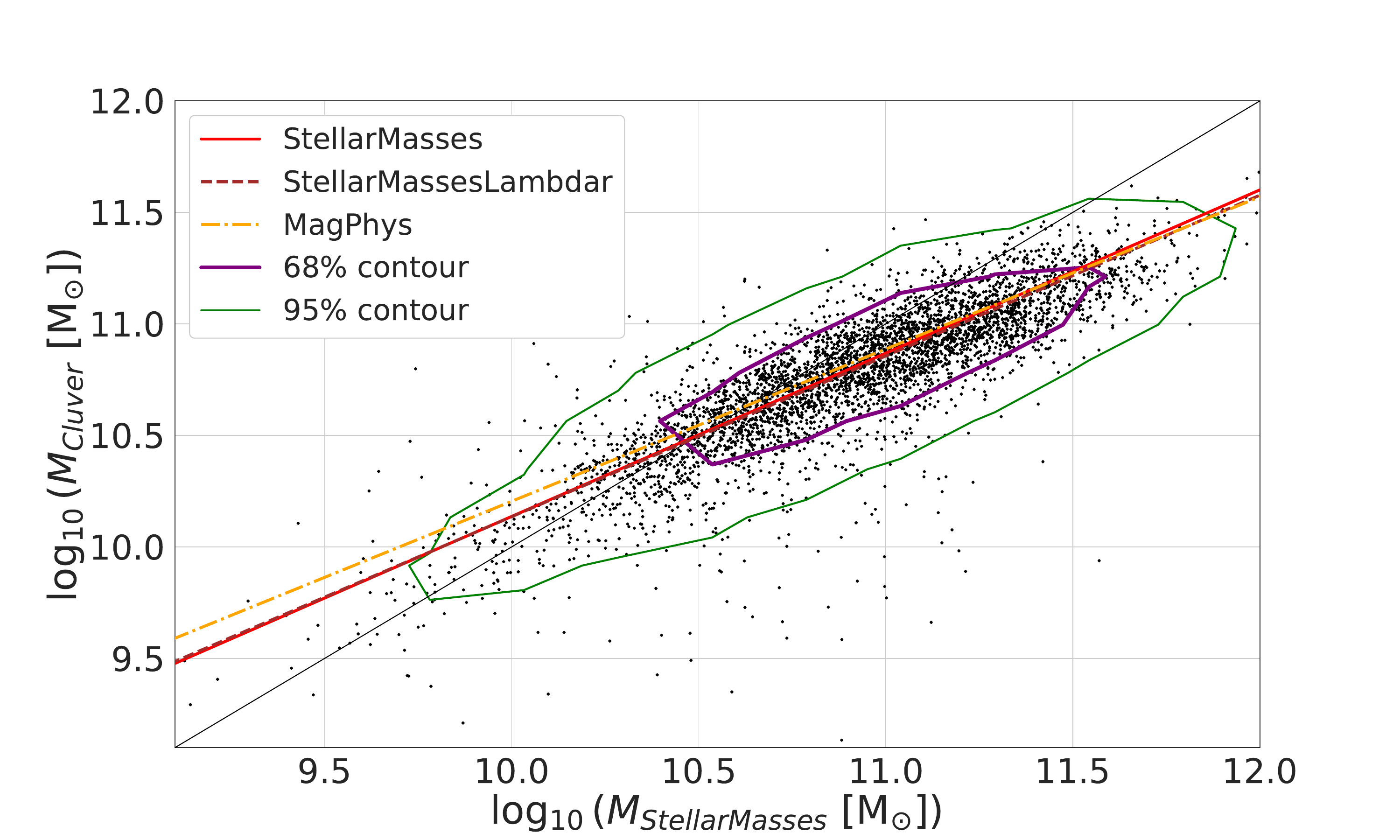}}\\
\subfloat{\includegraphics[width = 3.45in]{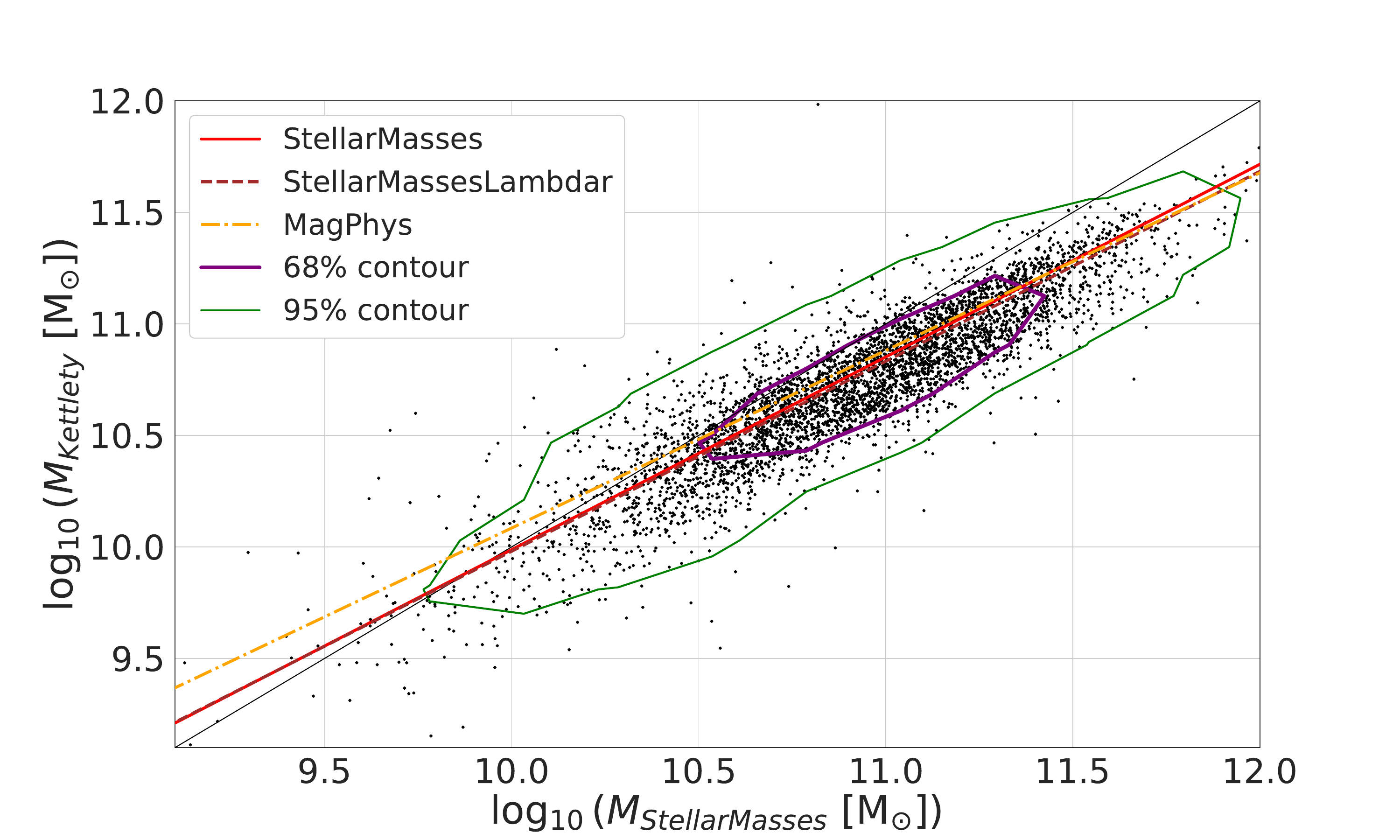}}
\subfloat{\includegraphics[width = 3.45in]{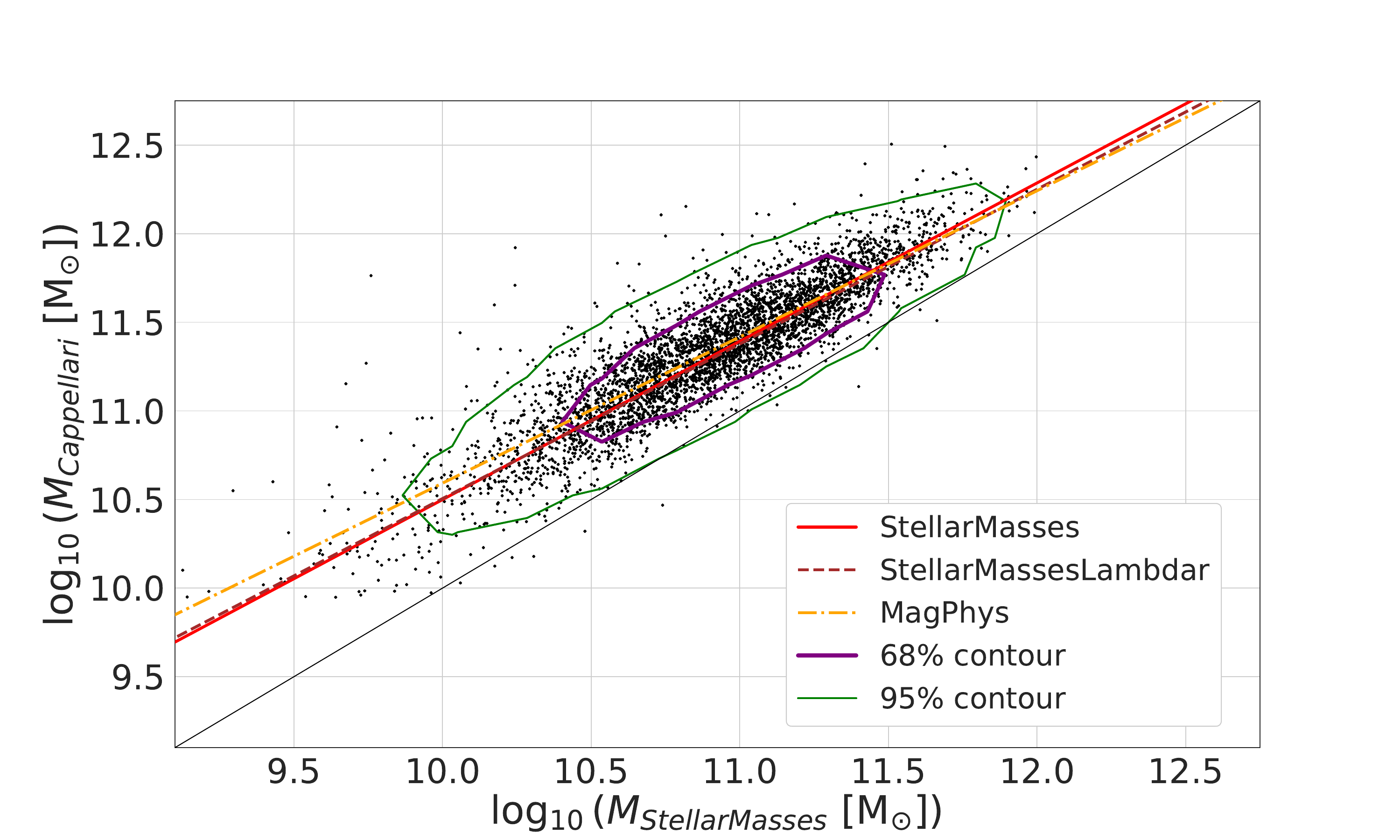}}
\caption{Log stellar masses from the four tested methods as functions of log stellar masses from the StellarMasses table of the GAMA catalogue. The purple contour outlines 68 per cent of the galaxies, and the green contour outlines 95 per cent. The black line denotes the ideal (identity) relationship. The red line shows the linear fit with the StellarMasses table, the brown dotted line with the StellarMassesLambdar table, and the orange dashed line with the MagPhys table.}
\label{fig:compare}
\end{figure*}

\begin{table*}
\caption{Results of $y=a  x + b$ linear fitting to the log stellar masses from the four tested methods ($y$) as the function of the log stellar masses ($x$) in the GAMA catalogue with the least square method.}
\label{tab:fitted_vals}
\begin{tabular}{cccccccccc}
\hline
                    & \multicolumn{3}{c|}{\textbf{a}}
                    &
                    & \multicolumn{3}{c|}{\textbf{b}}                                                                         \\ \cline{2-4} \cline{6-8}
                    & \textbf{StellarMasses} & \textbf{StellarMassesLambdar} & \textbf{MagPhys} && \textbf{StellarMasses} & \textbf{StellarMassesLambdar} & \textbf{MagPhys}       \\ \hline
\textbf{Jarrett}   & 0.713$\pm$0.002        & 0.701$\pm$0.002              & 0.660$\pm$0.002  && 2.82$\pm$0.02          & 2.93$\pm$0.02        & 3.41$\pm$0.02 \\ 
\textbf{Cluver}     & 0.732$\pm$0.001        & 0.721$\pm$0.001              & 0.68248$\pm$0.0009  && 2.81$\pm$0.01          & 2.93$\pm$0.01        & 3.38$\pm$0.01 \\ 
\textbf{Kettlety}   & 0.864$\pm$0.004        & 0.852$\pm$0.004               & 0.796$\pm$0.003  && 1.35$\pm$0.04        & 1.46$\pm$0.04        & 2.12$\pm$0.04 \\ 
\textbf{Cappellari} & 0.893$\pm$0.003     & 0.873$\pm$0.002              & 0.825$\pm$0.002  && 1.56$\pm$0.03          & 1.78$\pm$0.03        & 2.34$\pm$0.03 \\ \hline
\end{tabular}
\end{table*}

\begin{figure*}
\subfloat{\includegraphics[width = 3.45in]{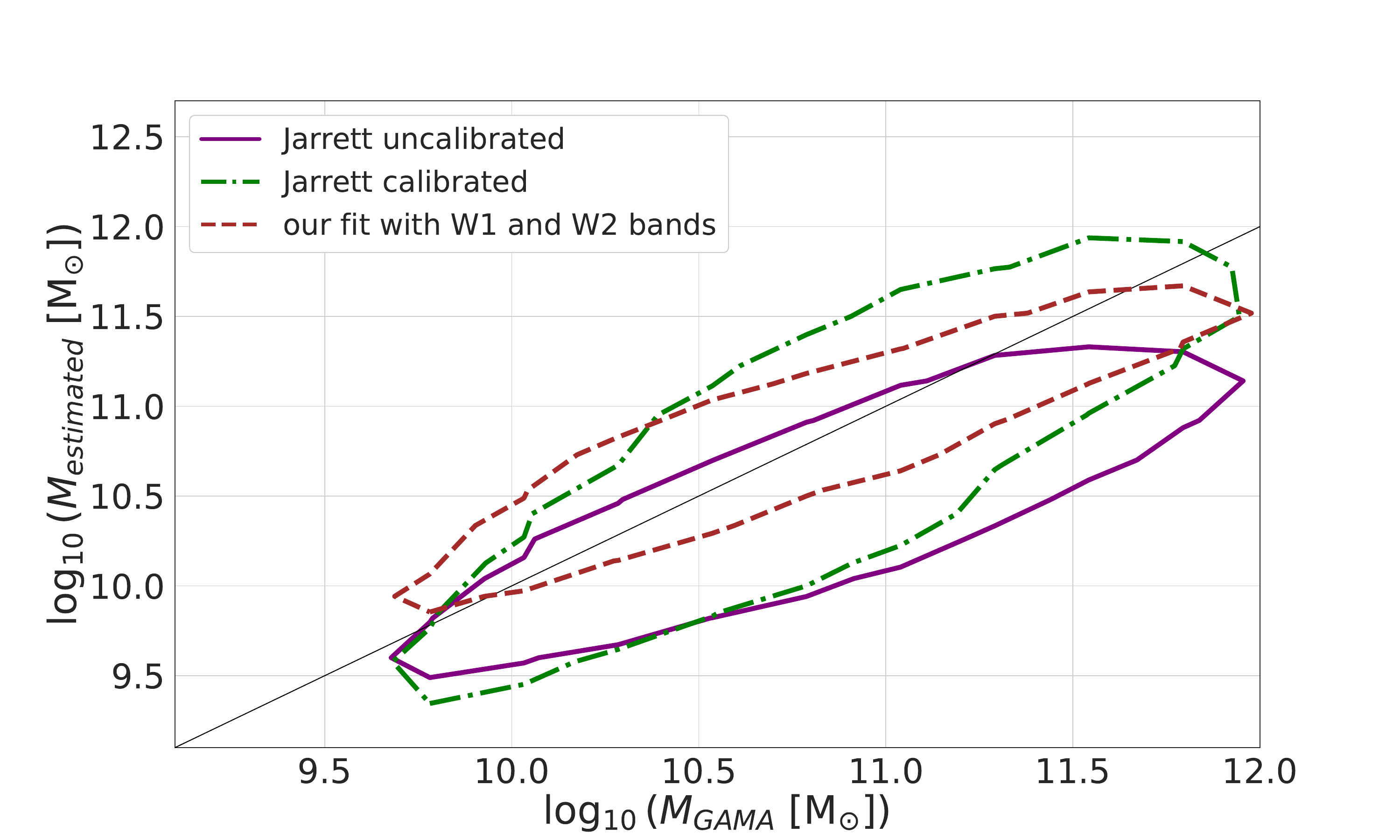}}
\subfloat{\includegraphics[width = 3.45in]{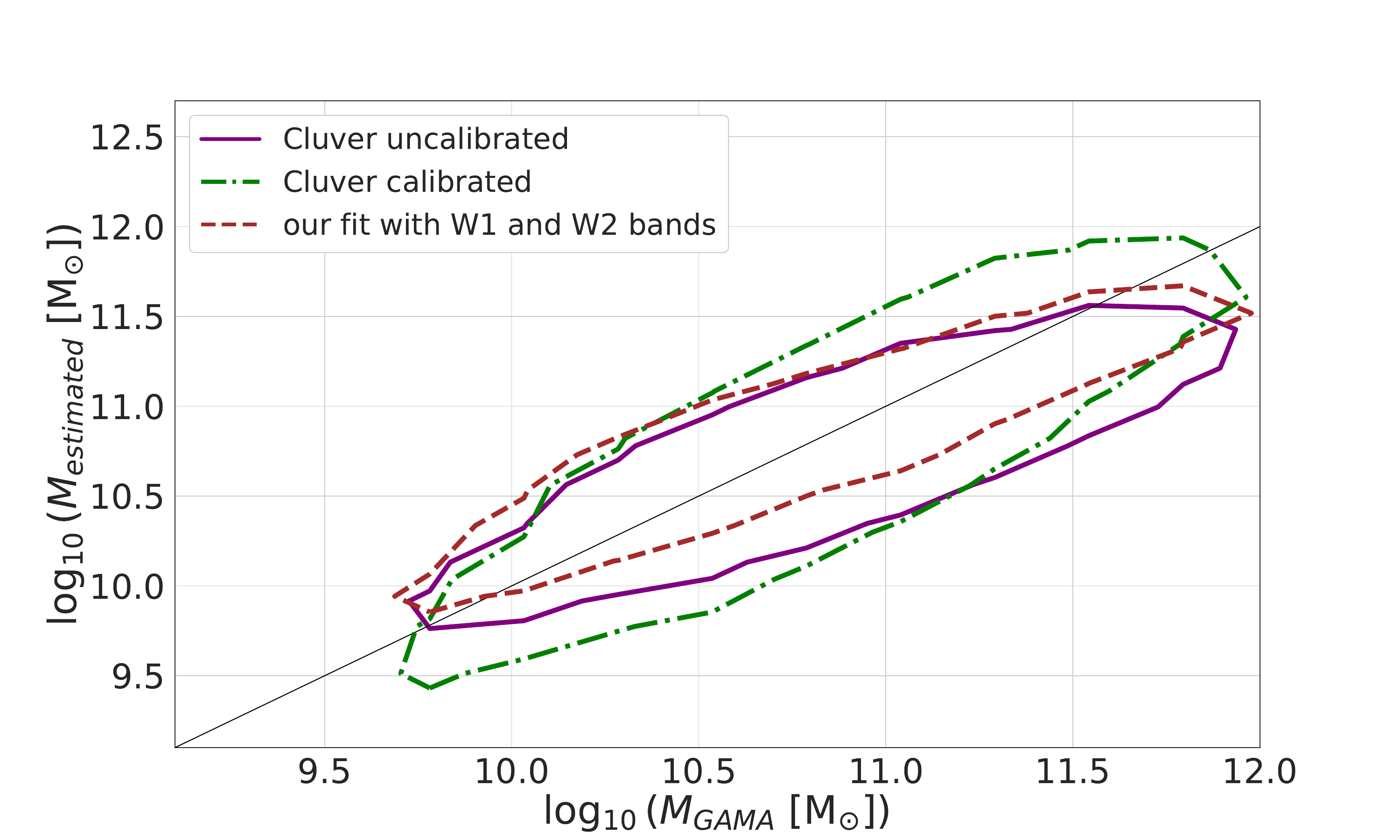}}\\
\subfloat{\includegraphics[width = 3.45in]{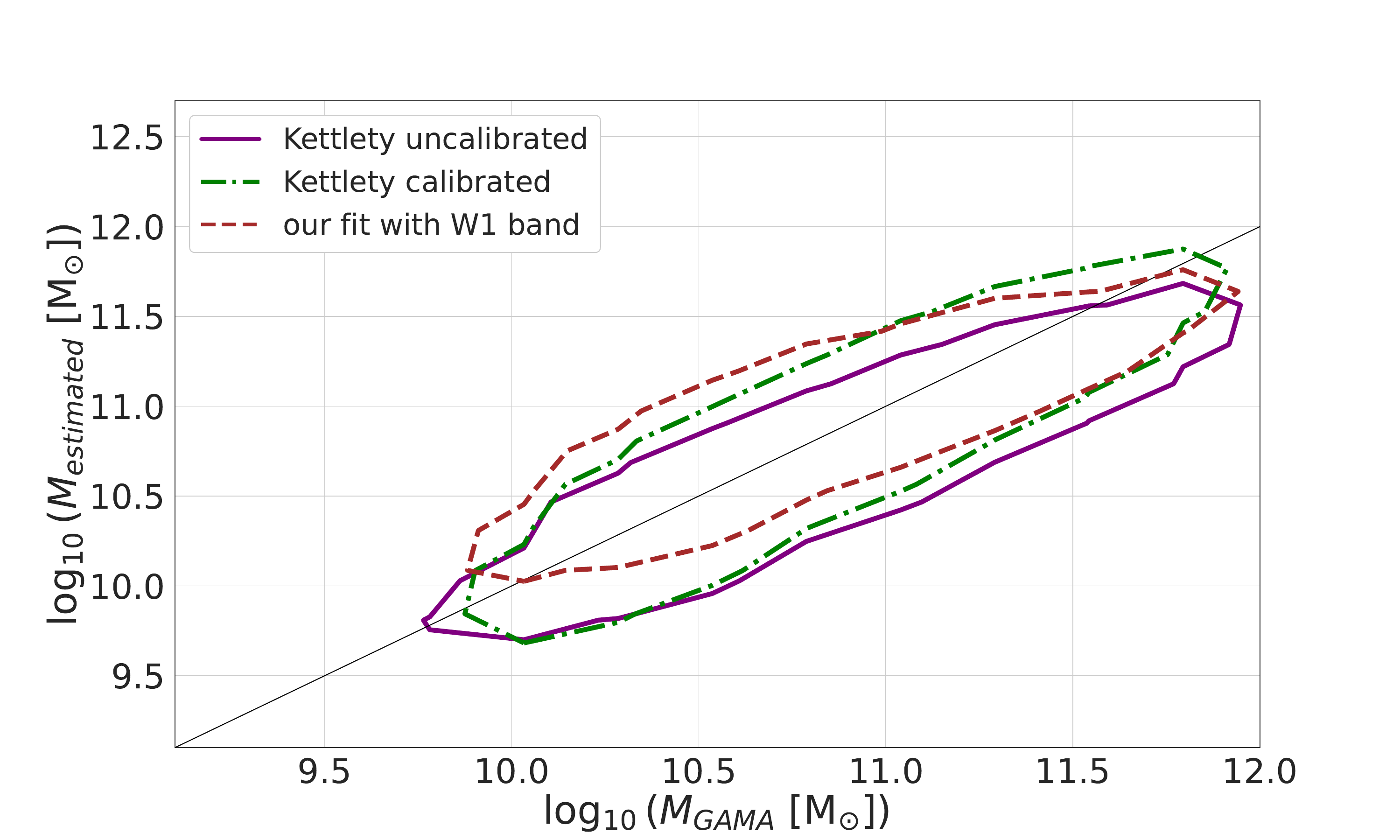}} 
\subfloat{\includegraphics[width = 3.45in]{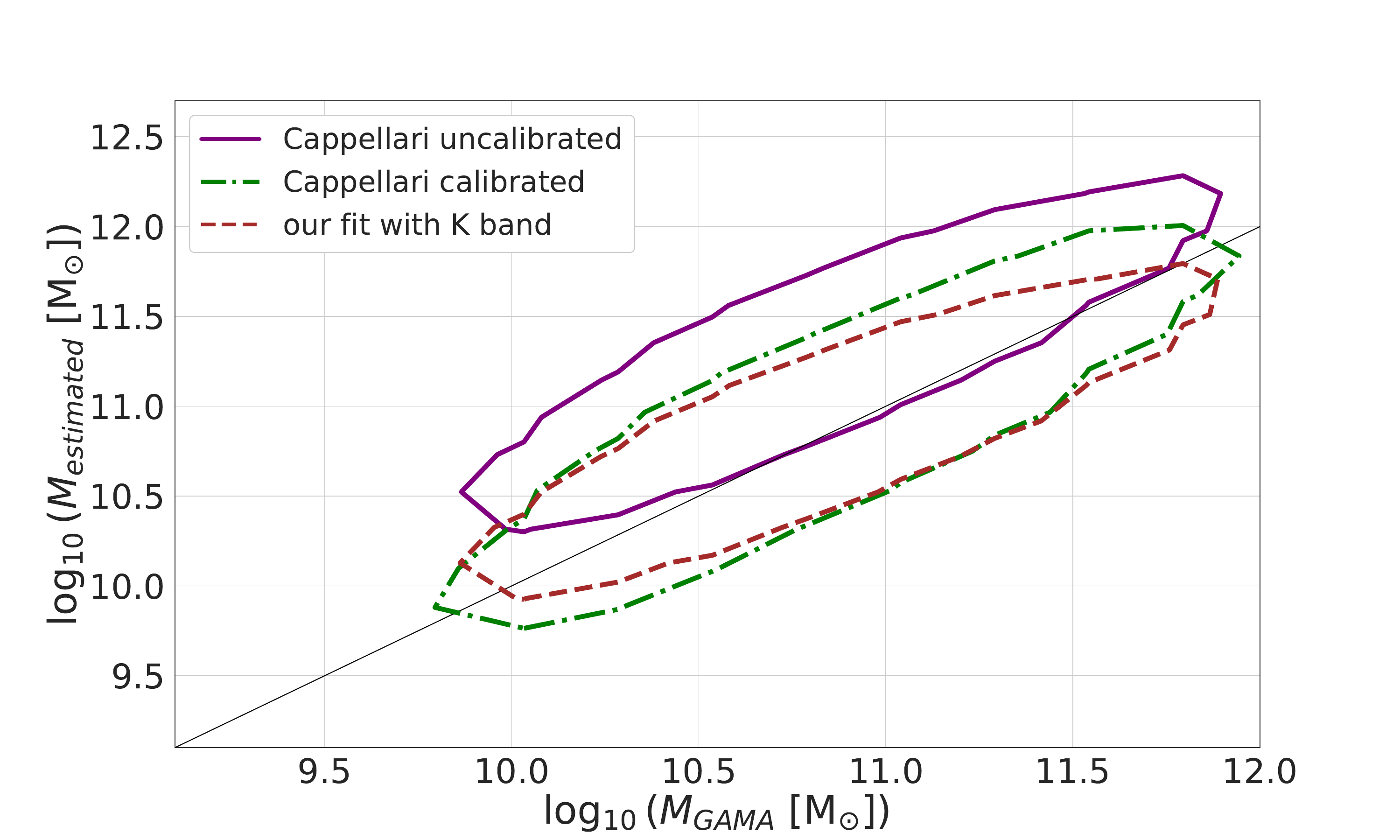}} 
\caption{Calibrating the results of the tested methods according to the linear relationship using the parameters in Table$~$\ref{tab:fitted_vals} for the StellarMasses values. The contour lines represent the area with 95 per cent of the estimated stellar masses as a function of the reference GAMA stellar masses from the StellarMasses table. The purple lines show the original (uncalibrated) results, the green dash-dot contours indicate the calibrated values, and the brown dashed contours represent the estimates based on our fit.}
\label{fig:shifted}
\end{figure*}

In Fig.$~$\ref{fig:compare}, we show the $\log_{10} M_\ast$ values from the four tested methods as functions of the StellarMasses $\log_{10} M_\ast$ values, with each curve representing the linear relationship between the tested method and the corresponding GAMA tables. These plots can be compared to Fig.$~$\ref{fig:GAMA}, which displays the relationship between the $\log_{10} M_\ast$ values of the different GAMA tables. We used the least squares method to obtain the best-fitting parameters, listed in Table$~$\ref{tab:fitted_vals}. 
Here, we can observe the same trends as in Fig.$~$\ref{fig:fraction}. 
The first three columns (labelled as $\sigma$) in Table$~$\ref{tab:chi} report the standard deviations of the linear fit for the four tested SME methods and the three GAMA tables. The last three columns of Table$~$\ref{tab:chi} show the root mean squared error introduced in equation$~$(\ref{eq:chi2}).

We can use the fitted values from Table$~$\ref{tab:fitted_vals} to calibrate the estimated stellar masses to the reference GAMA values. Fig.$~$\ref{fig:shifted} illustrates this calibration for the MagPhys values, showing only the 95 per cent contours and presenting the results based on the parameters we fitted of the tested methods.
We also fitted the relationships for the tested methods, which are represented by the brown dashed contours.
This figure demonstrates that the calibrated estimates align more closely with the ideal case (black lines) than the original values. The estimates based on our fits have similar properties in the case of the methods based on the WISE magnitudes, while they align much closer to the black line for the method based on the \textit{k\_m\_ext} magnitude. Overall, the calibrated values better match the ideal case than estimates from our fit.

\begin{table*}
\caption{The first three columns of the table (labelled as $\sigma$) show the standard deviations during the linear fit. The last three columns indicate the root mean squared error ($D$ values, see equation$~$\ref{eq:chi2}) of the four tested SME methods and the three reference stellar mass tables of the GAMA catalogue.}
\label{tab:chi}
\begin{tabular}{ccccccccc}
\hline
                    & \multicolumn{3}{c}{$\mathbf{\sigma}$} && \multicolumn{3}{c}{$\mathbf{D}$} \\ \cline{2-4} \cline{6-8}
                    & \textbf{StellarMasses} & \textbf{StellarMassesLambdar} & \textbf{MagPhys} && \textbf{StellarMasses} & \textbf{StellarMassesLambdar} & \textbf{MagPhys} \\ \hline
\textbf{Jarrett}    & 0.1971   & 0.1967 & 0.2079 && 0.4427                   & 0.4592                                                                     & 0.4303          \\
\textbf{Cluver}     & 0.1840 & 0.1839 & 0.1953 && 0.2622                   & 0.2749                                                                     & 0.2700            \\
\textbf{Kettlety}   & 0.1667 & 0.1681 &  0.1774 & & 0.2590                   & 0.2745                                                                    & 0.2553             \\
\textbf{Cappellari} & 0.1761   & 0.1802 & 0.1891 & & 0.4731                 & 0.4603                                                                    & 0.5120       \\ \hline
\end{tabular}
\end{table*}

\subsection{Ranking host candidates with their stellar masses}\label{sec:weight}
To assess the potential of these SME methods in identifying host galaxies of GW events, we studied the case of GW170817 \citep{GW170817}, the only GW event to date with identified host \citep{bright_siren}. We selected galaxies from the GLADE+ catalogue that were located within the 90 per cent localisation volume based on the initial BAYESTAR and the preliminary LALInference skymaps\footnote{\url{https://dcc.ligo.org/LIGO-G1701985/public}}
of GW170817. Table$~$\ref{tab:NGC4993} summarizes the results. $\mathit{N}$ indicates the number of galaxies in the localisation volume, $\mathit{r_\text{pos}}$ is the rank based on the localisation alone, and $\mathit{r_\text{tot}}$ is the rank when including the $M_\ast$ using equation$~$\ref{eq:weight2}. Using the initial BAYESTAR skymap, the host galaxy ranked ninth based solely on localisation but moved to the first place when stellar mass was considered, regardless of the estimation method (Jarrett's, Cluver's, Kettlety's or Cappellari's) or  ranking approach (Ducoin's or Artale's). For the preliminary LALInference skymap, $\mathit{r_\text{pos}}$ was 27, improving to 16 with Ducoin's approach using WISE-based $M_\ast$ estimates and to 21 with $M_\ast$ from Cappellari's method. With Artale's ranking approach, the $\mathit{r_\text{tot}}$ improved to 16, 16, 18, and 22 for Jarrett's, Cluver's, Kettlety's and Cappellari's SME methods, respectively. It is worth noting that the WISE$\times$SuperCOSMOS catalogue contains more galaxies than the 2MASS catalogue, resulting in more galaxies having stellar mass estimates from WISE-based methods compared to Cappellari's method. For the BAYESTAR skymap, 50 galaxies have $M_\ast$ estimates from WISE-based methods, while only 14 have estimates from Cappellari's method. Similarly, for the LALInference skymap, these numbers are 48 and 13, respectively. Since some galaxies lack stellar mass estimates, using the modified ranking formula in equation \ref{eq:weight2} is necessary, as the original definition of $G_\text{tot}$ (equation \ref{eq:weight}) does not account for missing values.
These tests already show the potential of considering the stellar mass from the tested SME methods in the host candidate ranking.

\begin{table*}
\caption{Ranking of NGC4993, the host galaxy of GW170817, using the initial BAYESTAR and the preliminary LALInference skymaps. $\mathit{N}$ denotes the number of galaxies, $\mathit{r_\text{pos}}$ is the rank based on the localisation volume alone, and $\mathit{r_\text{tot}}$ gives the rank when including the $M_\ast$ estimated with the tested methods  (section$~$\ref{sec:method:rank}). For the BAYESTAR skymap, both Ducoin's and Artale's approaches gave the same ranks for the true host. Including stellar mass significantly improved the rank of the host in all cases.}
\label{tab:NGC4993}
\begin{tabular}{ccccccccccc}
\hline
\textbf{}                              &        &  & \multirow{2}{*}{$\mathbf{N}$} &  & \multirow{2}{*}{$\mathbf{r_\text{pos}}$} &  & \multicolumn{4}{c}{$\mathbf{r_\text{tot}}$} \\ \cline{8-11} 
                                       &        &  &                               &  &                                          &  & Jarrett  & Cluver  & Kettlety  & Cappellari \\ \hline
\textbf{BAYESTAR}                      &        &  & 58                            &  & 9                                        &  & 1        & 1       & 1         & 1          \\
\\
\multirow{2}{*}{\textbf{LALInference}} & Ducoin &  & \multirow{2}{*}{56}           &  & \multirow{2}{*}{27}                      &  & 16       & 16      & 16        & 21          \\
                                       & Artale &  &                               &  &                                          &  & 16        & 16       & 18        & 22          \\ \hline
\end{tabular}
\end{table*}

\begin{figure}
    \centering
    \includegraphics[width=1\linewidth]{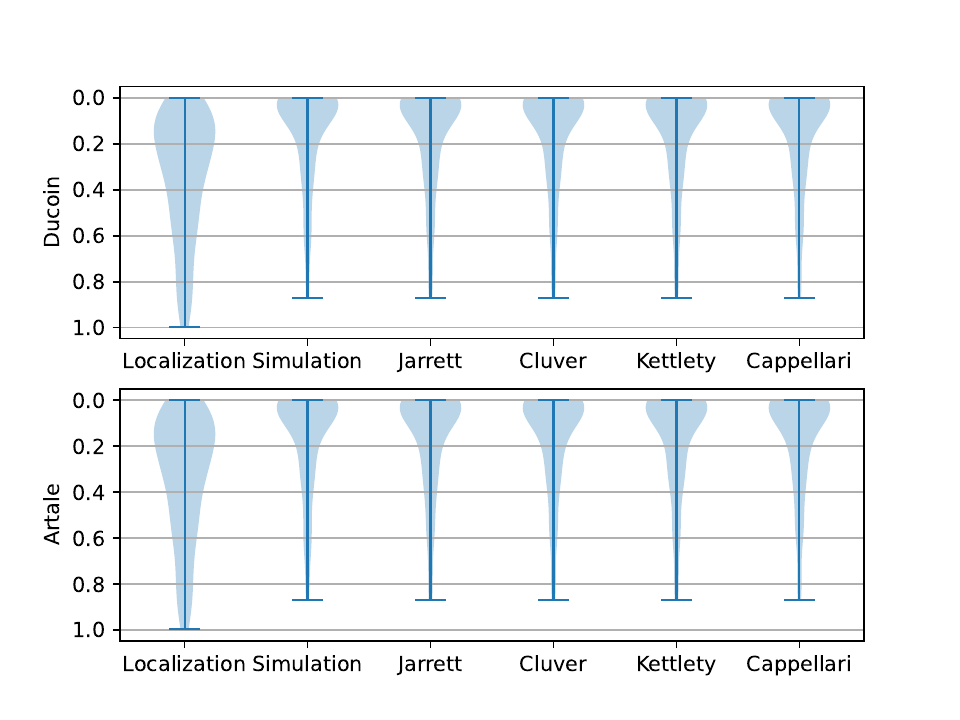} 
    \caption{Violin plots of the normalized ranks of the hosts for simulated BBHs: the host rank divided by the number of possible hosts per event. Each column represents different ranking approaches: using localisation only, and using localisation combined with stellar mass from the simulation (true values), Jarrett's, Cluver's, Kettlety's, and Cappellari's methods. The top plot shows results from Ducoin's approach, and the bottom plot from Artale's. The vertical lines represent the minimum and maximum values. The plots demonstrate that incorporating stellar mass improves the ranking, resulting in lower normalized ranks and better chances of identifying the true hosts.}
    \label{fig:rank}
\end{figure}

We used simulated data described in Section$~$\ref{sec:method:rank} to further investigate the potential benefits of incorporating stellar mass estimates in host galaxy ranking. To ensure that no unrealistic $M_\ast$ were included in the analysis, we excluded SMEs from the IR-based methods that fell outside the range $[10^6 ~M_\odot, 10^{13}~M_\odot]$. 
Fig.$~$\ref{fig:rank} shows the violin plots of the distributions of the normalized ranks (calculated as the rank from the ranking approach divided by the number of host candidates for each event) for the host galaxies of individual events. A lower ranking value corresponds to a higher placement of the host in the potential host galaxy list. The violin plots clearly show that accounting for the stellar mass improves the ranking.  This is also evident from the first rows of Tables$~$\ref{tab:rank_D}$~$and$~$\ref{tab:rank_A}, which present the fractions of cases where rankings that account for stellar mass ($\mathit{r_\text{tot}}$) outperform those based solely on localisation ($\mathit{r_\text{pos}}$). 

\begin{figure}
    \centering
    \includegraphics[width=1\linewidth]{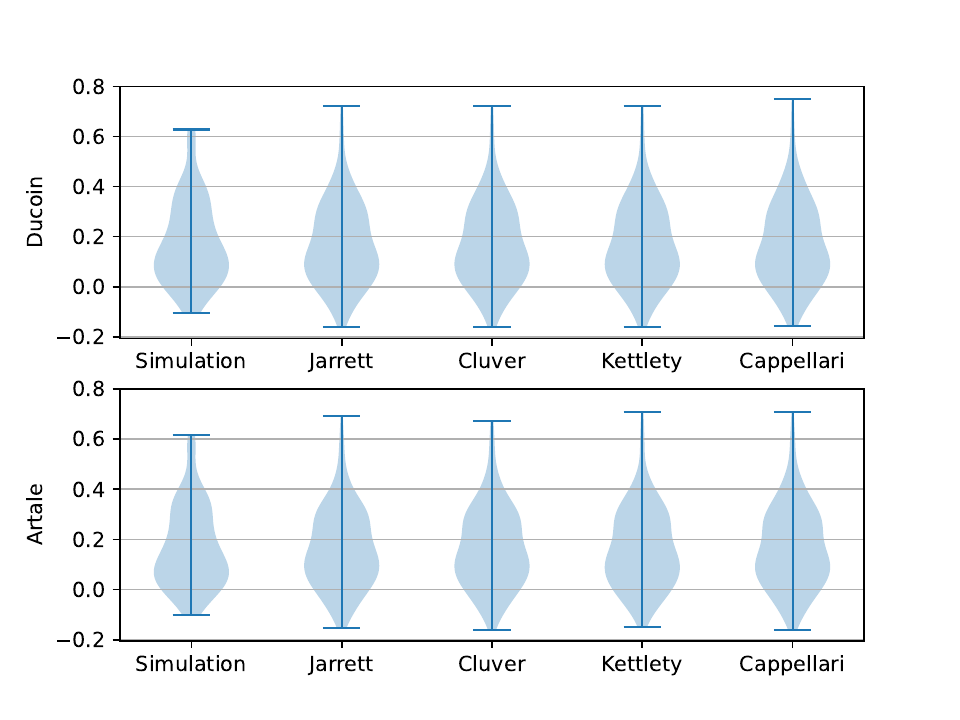} 
    \caption{Violin plots of normalized rank differences, as defined in equation$~$(\ref{eq:NRD}): the difference between the rank of the host based on localisation alone and the rank accounting for both localisation and stellar mass, divided by the number of hosts for each event. The columns represent rankings using stellar mass from the simulation (true value), Jarrett's, Cluver's, Kettlety's, and Cappellari's methods. The top plot shows results from Ducoin's approach, and the bottom plot from Artale's. The vertical lines represent the minimum and maximum values. The similar shapes of the violin plots indicate that stellar masses from the different methods lead to comparable improvements in rankings.}
    \label{fig:nrd}
\end{figure}

To determine the improvement in rankings based on the different stellar mass estimates, we defined the normalized rank difference (NRD) for event $i$ as
\begin{equation}\label{eq:NRD}
    (\text{NRD})_i = \frac{r_{\text{pos},i}-r_{\text{tot}, i}}{N_i},
\end{equation}
where $N_i$ is the number of possible host galaxies for event $i$. Fig.$~$\ref{fig:nrd} presents violin plots of normalized rank differences for the two approaches presented in Section$~$\ref{sec:method:rank} using five different stellar mass values: the true value from the simulation and those estimated with Cluver's, Jarrett's, Kettlety's and Cappellari's methods. The similarity in the distribution shapes indicates that rankings based on stellar masses from luminosity-based methods perform comparably to those based on the true stellar mass. 

To assess whether these results are specific to this simulation or represent a more general conclusion, we applied bootstrap resampling. We resampled the $\text{NRD}_i$ values 10 000 times and calculated the mean, standard deviation, and their uncertainties based on the 95 per cent confidence intervals. These values are presented in the second and third rows of Tables$~$\ref{tab:rank_D}$~$ and$~$\ref{tab:rank_A} for the approaches of \citet{Ducoin} and \citet{Artale_2020}, respectively. The mean and standard deviations of the NRDs with the different stellar mass estimates agree within the margin of error, indicating that, based on this simulation, the improvement from using the simulated $M_\ast$ value is not better than using stellar masses from the luminosity-based methods. Additionally, the small estimated uncertainties suggest that the results are not specific to our simulation. 

\begin{table*}
\caption{Ranking of host galaxies for the simulated events, incorporating different stellar mass estimates and using the ranking approach of \citet{Ducoin} described in Section$~$\ref{sec:method:rank}. These are compared to the host's rank solely based on localisation probability. The first row shows the percentage of cases where rankings with stellar masses outperform those based on localisation alone (a ranking is better if the host's rank is lower). The second row provides the means of the normalized rank differences (defined in equation$~$\ref{eq:NRD}), and the third row shows the standard deviations. Uncertainties for the second and third rows were calculated using bootstrap resampling.  For better readability, we combined the upper and lower uncertainty bounds where they were equal.}
\label{tab:rank_D}
\begin{tabular}{|c|c|c|c|c|c|}
\hline
                                   & \textbf{Simulated} & \textbf{Jarrett} & \textbf{Cluver} & \textbf{Kettlety}    & \textbf{Cappellari} \\ \hline
\textbf{Fraction [\%]} & 85                 & 84                 & 84               & 84                     & 84  
\\ 
\textbf{Mean NRD}                  & $0.17 \pm 0.03$    & $0.17 \pm 0.03$   & $0.17 \pm 0.03$  & $0.17 \pm 0.03$  &   $0.17 \pm 0.03$  \\
\textbf{STD NRD}                   & $0.17 \pm 0.02$     & $0.16 \pm 0.02$   & $0.16 \pm 0.02$       & $0.16 \pm 0.02$   &  $0.17 \pm 0.02$  \\ \hline
\end{tabular}         
\end{table*}

\begin{table*}
\caption{Same as Table$~$\ref{tab:rank_D}, but using the ranking approach of \citet{Artale_2020}.}
\label{tab:rank_A}
\begin{tabular}{|c|c|c|c|c|c|}
\hline
                                   & \textbf{Simulated} & \textbf{Jarrett} & \textbf{Cluver}      & \textbf{Kettlety}    & \textbf{Cappellari}  \\ \hline
\textbf{Fraction [\%]} & 84                   & 84               & 85                   & 83                     & 85                     \\ 
\textbf{Mean NRD}                  & $0.17 \pm 0.03$     & $0.16 \pm 0.03$   &  $0.17 \pm 0.03$       & $0.16 \pm 0.03$      & $0.17 \pm 0.03$       \\ 
\textbf{STD NRD}                   & $0.16 \pm 0.02$     & $0.16 \pm 0.03$   & $0.16 \pm 0.02$  & $0.16 \pm 0.02$  & $0.16 \pm 0.02$ \\ \hline
\end{tabular}         
\end{table*}

\section{Conclusions}\label{sec:Conclusions}
We compared the results of four luminosity-based SME methods that use IR band magnitudes with the more accurate stellar masses from GAMA DR3. Based on our findings, we assess the best SME method for GW astronomy. Fig.$~$\ref{fig:corr} highlights that Cappellari's method shows the strongest correlation with the GAMA stellar masses, and Table$~$\ref{tab:fitted_vals} indicates that the slope of its fitted linear slope is the closest to 1. 
Additionally, Cappellari's method yields the smallest estimated uncertainties for the entire GLADE+ sample (see Table$~$\ref{tab:err}), although it systematically overestimates the stellar masses. Kettlety's method achieves the smallest average standard deviation in linear fitting  and the lowest root mean squared error on average, as shown in Table$~$\ref{tab:chi}. We observed that the methods using WISE magnitudes (Jarrett's, Cluver's, and Kettlety's) tend to underestimate the $M_\ast$ values for massive galaxies ($M_\ast > 10^{10} ~ \text{M}_\odot$), whereas this trend is less pronounced for Cappellari's method.

Next, we ranked the galaxies within the localisation area of simulated BBH events based on their localisation probability and stellar masses, either from the simulated galaxy catalogue from which we sampled the host galaxies or estimated using the tested methods (see Section$~$\ref{sec:weight}). We implemented two ranking approaches \citep{Ducoin, Artale_2020} and demonstrated that considering stellar mass improved the ranking in over 80 per cent of cases compared to using localisation alone.
The improvements are consistent within the error limits, whether we use the stellar masses from the simulated catalogue or those estimated by the tested methods (see Tables$~$\ref{tab:rank_D}$~$and$~$\ref{tab:rank_A}). Therefore, based on these tests, the IR-based methods perform similarly effectively for ranking.  As discussed, Kettlety’s and Cappellari’s methods can be interpreted as ranking by W1 or K-band luminosities in the context of Ducoin’s approach. Our results show that these methods perform just as well as the other stellar mass estimates. These findings confirm that while including stellar mass weighting is important, the specific choice of SME method is less critical, as different methods yield comparable ranking improvements. This can be explained by the variation between different stellar mass estimates being small compared to the overall range of stellar masses in the galaxy population (see, e.g., Fig.$~$\ref{fig:compare}). Furthermore, the consistency between rankings from Ducoin’s and Artale’s approaches suggests that deviations from a strictly linear stellar mass–host probability relationship do not significantly impact ranking effectiveness.

 The effectiveness of stellar mass weighting in host galaxy ranking arises from the large variation in $M_\ast$ across the galaxy population. Since stellar mass estimates span several orders of magnitude, small uncertainties in $\log_{10} M_\ast$ have little impact on ranking results, as relative differences between galaxies remain significant. This remains true although modeling uncertainties are a greater concern than statistical or photometric errors in stellar mass estimates, as discussed in Section$~$\ref{sec:IRmethods} and illustrated in Fig.$~$\ref{fig:corr}.

 Our results suggest that simpler, luminosity-based SME methods can be as effective as more complex ones for GW host identification, meaning that practical considerations—such as data availability, survey depth, and computational efficiency—can play a key role in their selection. Methods that rely on fewer data inputs (e.g., Kettlety’s and Cappellari’s) and are computationally less expensive can serve as viable alternatives for ranking CBC host galaxies when more robust approaches are unavailable or impractical due to computational constraints.
Since WISE provides deeper infrared coverage than 2MASS, WISE-based SME methods offer a more complete galaxy sample. Additionally, deeper surveys such as the DESI Legacy Imaging Survey \citep{DESI} and simple optical color–$M_\ast$/L relations \citep[e.g.,][]{Bell2000, Taylor_11} may provide promising alternatives for future work.

\section*{Acknowledgements}
This publication makes use of data products from the Wide-field Infrared Survey Explorer, which is a joint project of the University of California, Los Angeles, and the Jet Propulsion Laboratory/California Institute of Technology, and NEOWISE, which is a project of the Jet Propulsion Laboratory/California Institute of Technology. WISE and NEOWISE are funded by the National Aeronautics and Space Administration
This publication makes use of data products from the Wide-field Infrared Survey Explorer, which is a joint project of the University of California, Los Angeles, and the Jet Propulsion Laboratory/California Institute of Technology, funded by the National Aeronautics and Space Administration.

Data used in this work was generated using Swinburne University's Theoretical Astrophysical Observatory (TAO). TAO is
part of the Australian All-Sky Virtual Observatory (ASVO) and is freely accessible at \url{https://tao.asvo.org.au/tao/}.
The Millennium Simulation was carried out by the Virgo Supercomputing Consortium at the Computing Centre of the Max
Plank Society in Garching. It is publicly available at \url{http://www.mpa-garching.mpg.de/Millennium/}.
The Semi-Analytic Galaxy Evolution (SAGE) model used in this work is a publicly available codebase that runs on the dark
matter halo trees of a cosmological N-body simulation. It is available for download at
\url{https://github.com/darrencroton/sage}.

The authors acknowledge the use of OpenAI's ChatGPT for assistance in refining the language and improving the clarity of the manuscript.

The authors are grateful for computational resources provided by the LIGO Laboratory and supported by  National Science Foundation's grant Nos. PHY-0757058 and PHY-0823459.
This project has received funding from the HUN-REN Hungarian Research Network and was also supported by the NKFIH excellence grant TKP2021-NKTA-64.

We sincerely thank the referee, Edward Taylor, for his valuable comments and suggestions, which have significantly contributed to improving the clarity and robustness of this work.

\section*{Data Availability}
The data underlying this article are available in Zenodo, at \url{https://doi.org/10.5281/zenodo.15054067}.

\appendix

\section{Differences of stellar mass estimates in redshift range z < 0.15}\label{app:diff}
\begin{figure*}
\subfloat{\includegraphics[width = 3.45in]{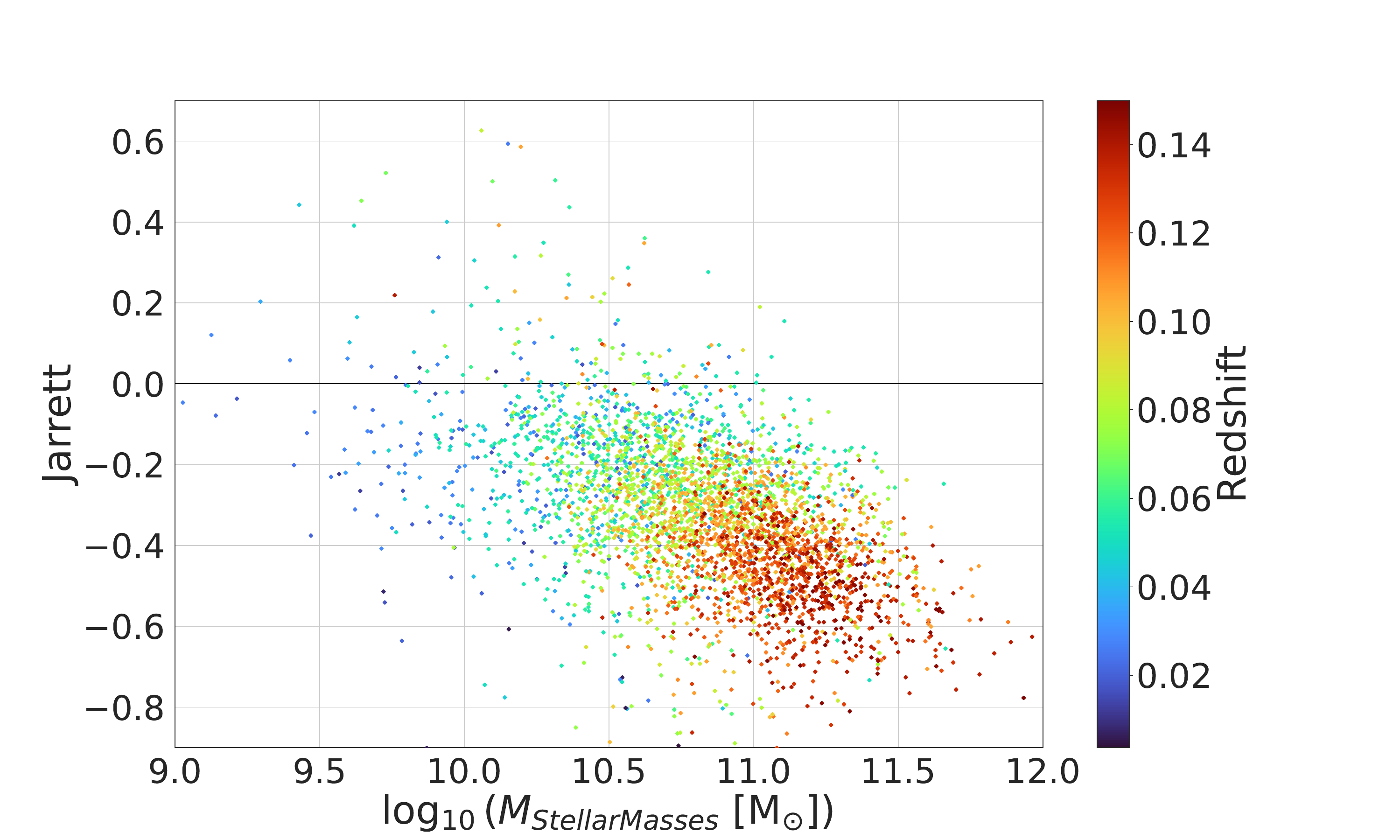}}
\subfloat{\includegraphics[width = 3.45in]{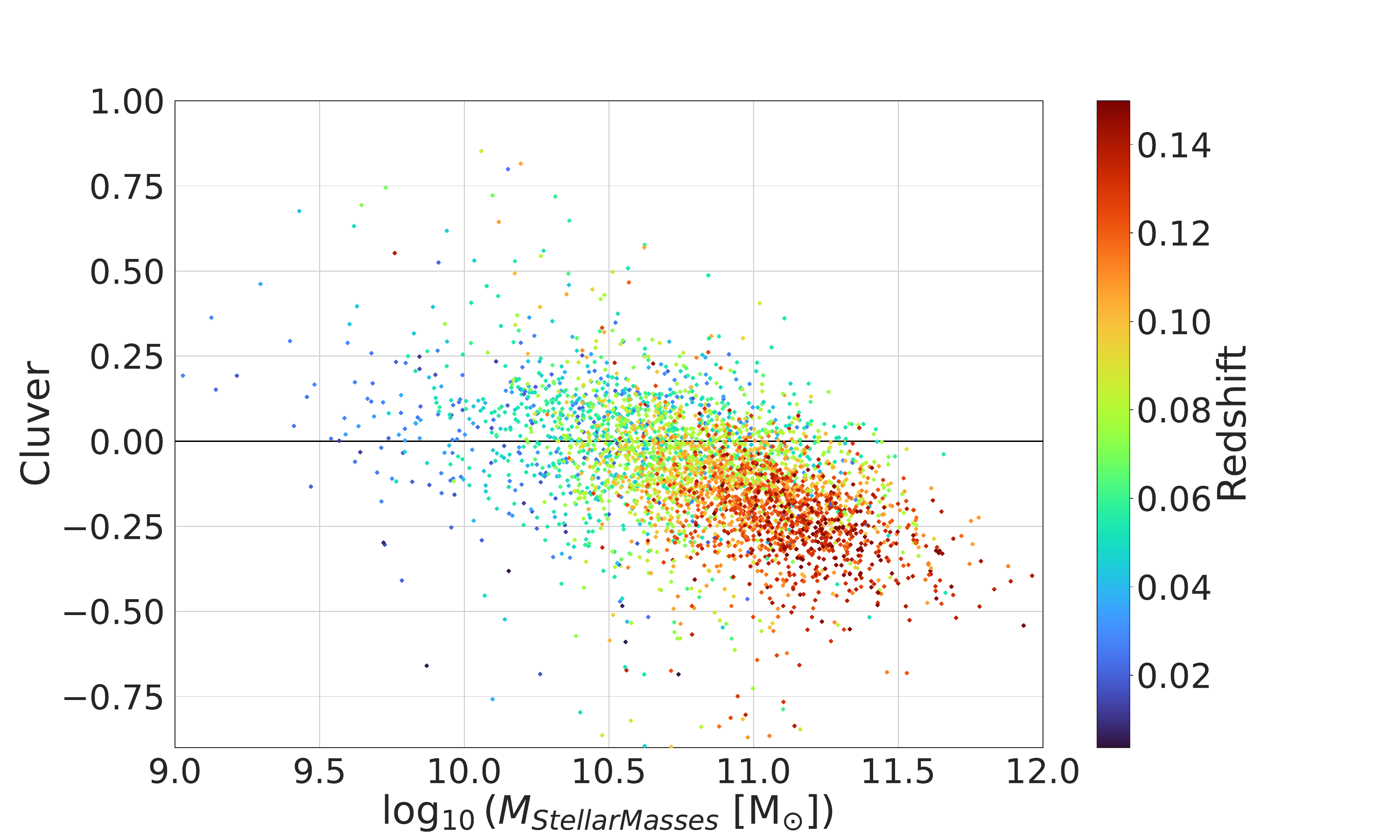}}\\
\subfloat{\includegraphics[width = 3.45in]{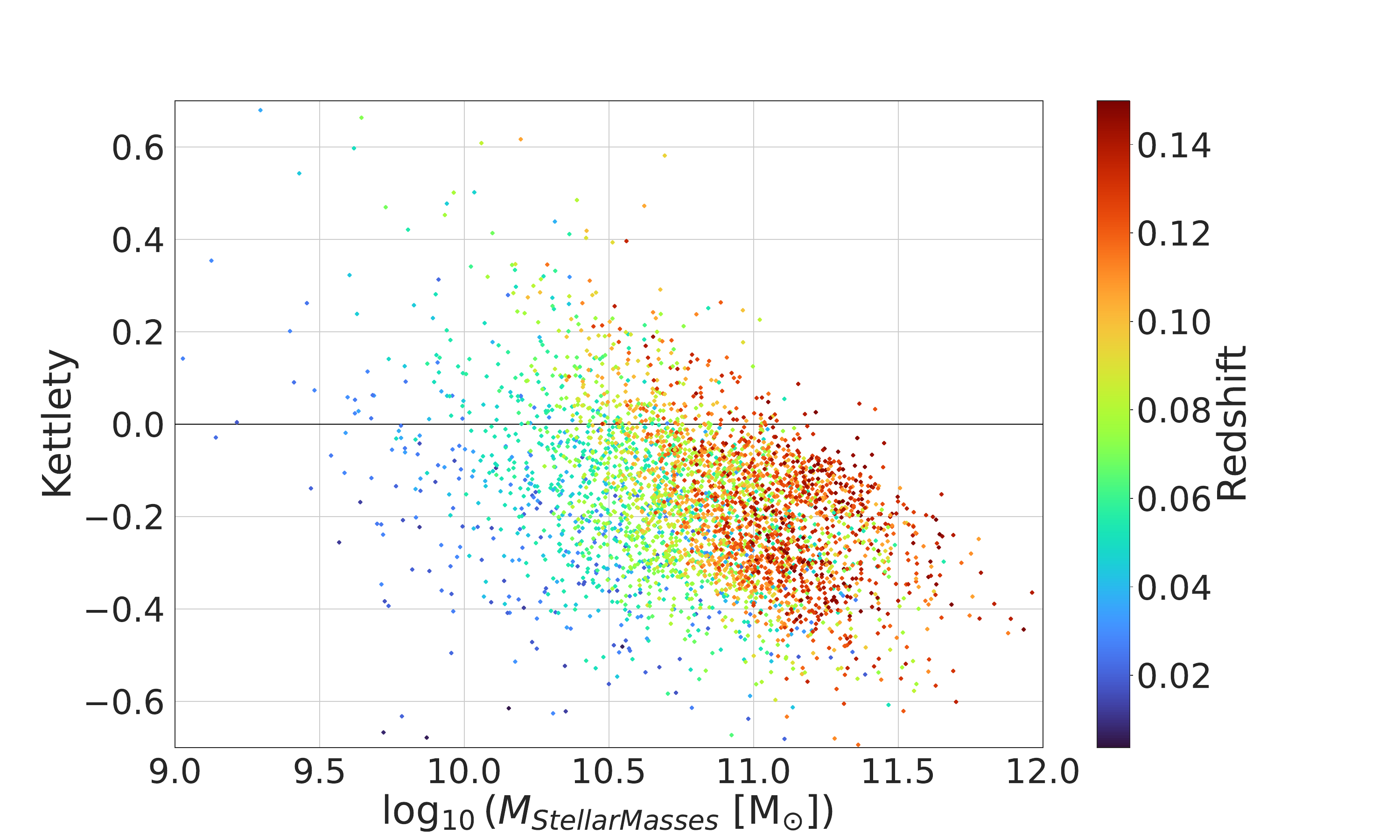}}
\subfloat{\includegraphics[width = 3.45in]{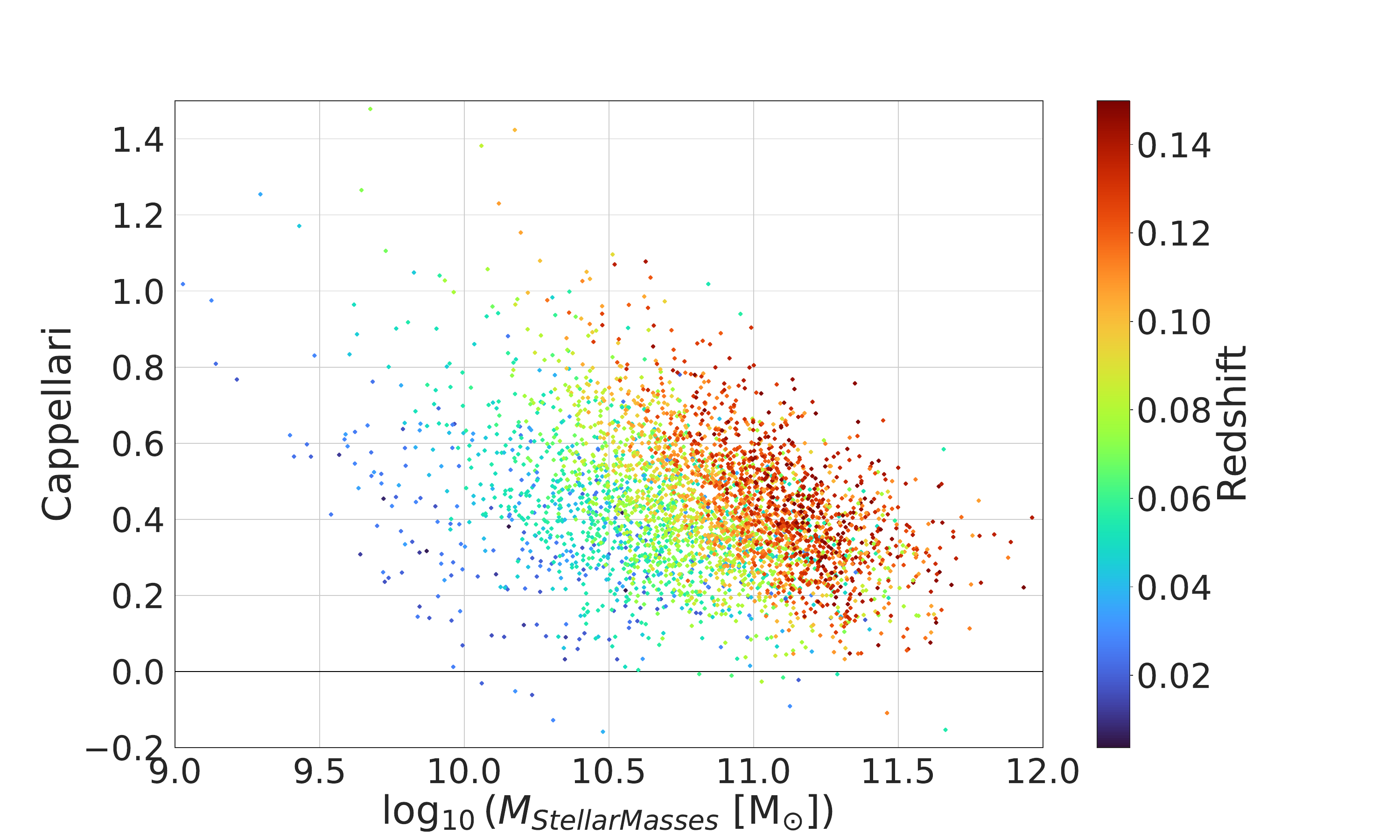}} 
\caption{Log difference between stellar masses derived from the tested methods and the StellarMasses values from the GAMA catalogue ($\log_{10}(M_\text{tested method}/M_\text{StellarMasses})$), plotted against the StellarMasses values for galaxies with redshift $z < 0.15$. The $y$-axis labels indicate the tested methods, and the colour code represents the redshifts of galaxies in the sample. Compare this figure with Fig.$~$\ref{fig:fraction}.}
\label{fig:fraction_15}
\end{figure*}

In this appendix, we present the differences between the stellar masses obtained from the tested methods and the GAMA values, which we treat as the actual stellar masses for galaxies with redshifts $z < 0.15$, because the stellar masses of \citet{Taylor_11} are best constrained in these cases. A total of 3,632 galaxies (about 85 per cent) in our sample meet this criterion. 
Fig.$~$\ref{fig:fraction_15}  shows these differences as a function of the $\log_{10} M_\ast$ values from the StellarMasses table of the GAMA catalogue.  Here, we observe similar trends to those in Fig.$~$\ref{fig:fraction}, which shows the differences for the entire sample. The differences between the results from Cluver's method and the StellarMasses table of the GAMA catalogue resembles fig.$~$8 of \citet{Cluver_2014}.

\section{Comparing stellar masses of GAMA tables}\label{app:B}
\begin{figure}
\subfloat{\includegraphics[width = 3.3in]{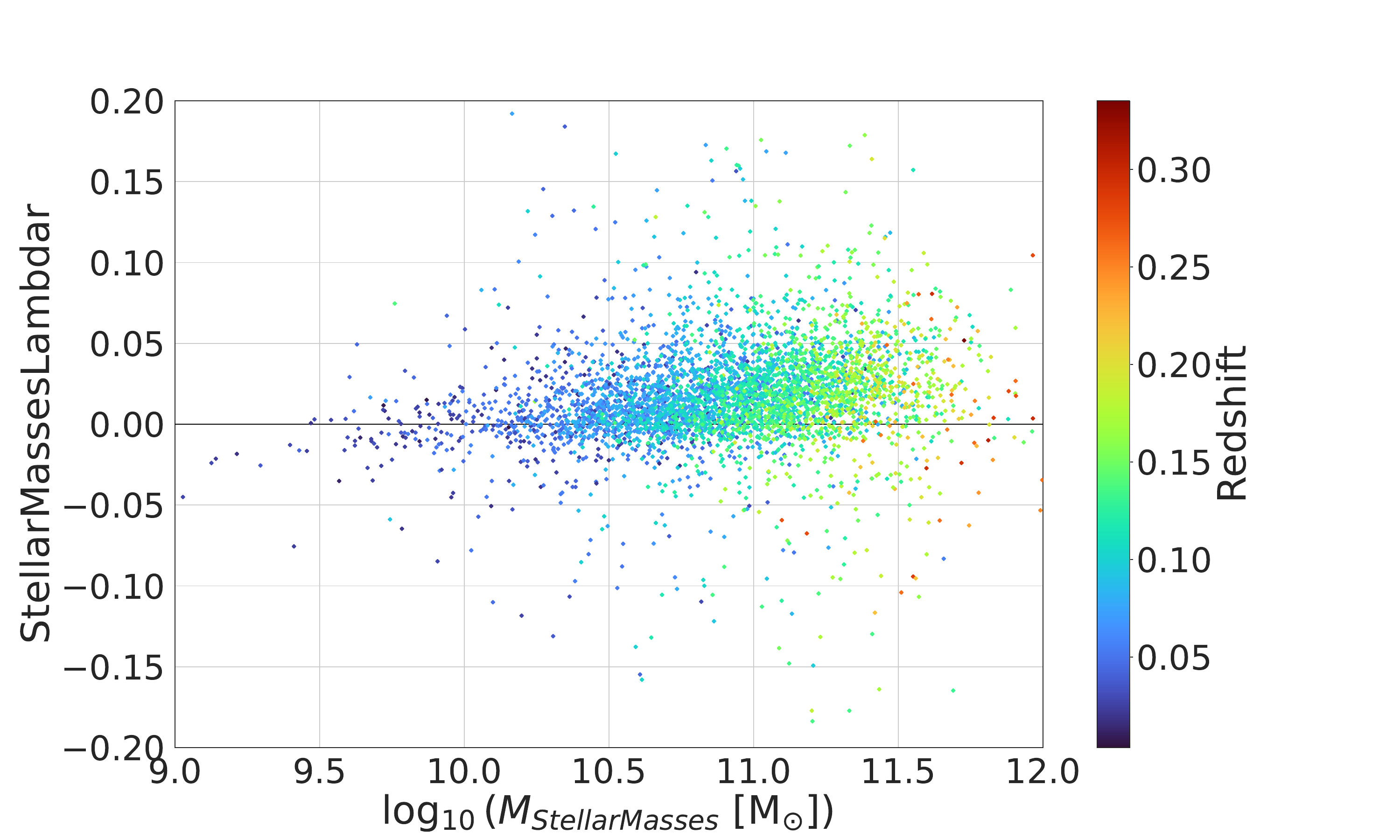}}\\
\subfloat{\includegraphics[width = 3.3in]{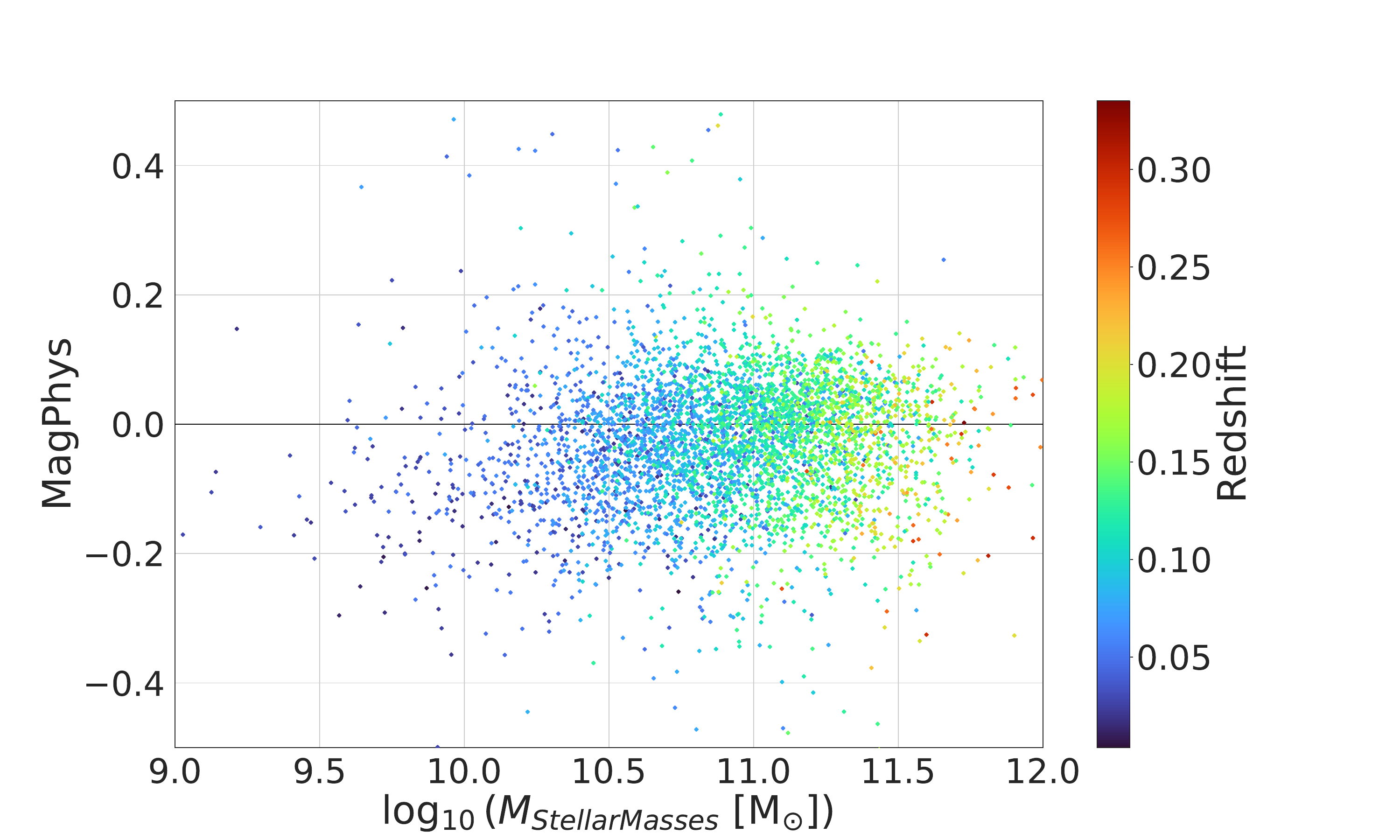}}\\
\subfloat{\includegraphics[width = 3.3in]{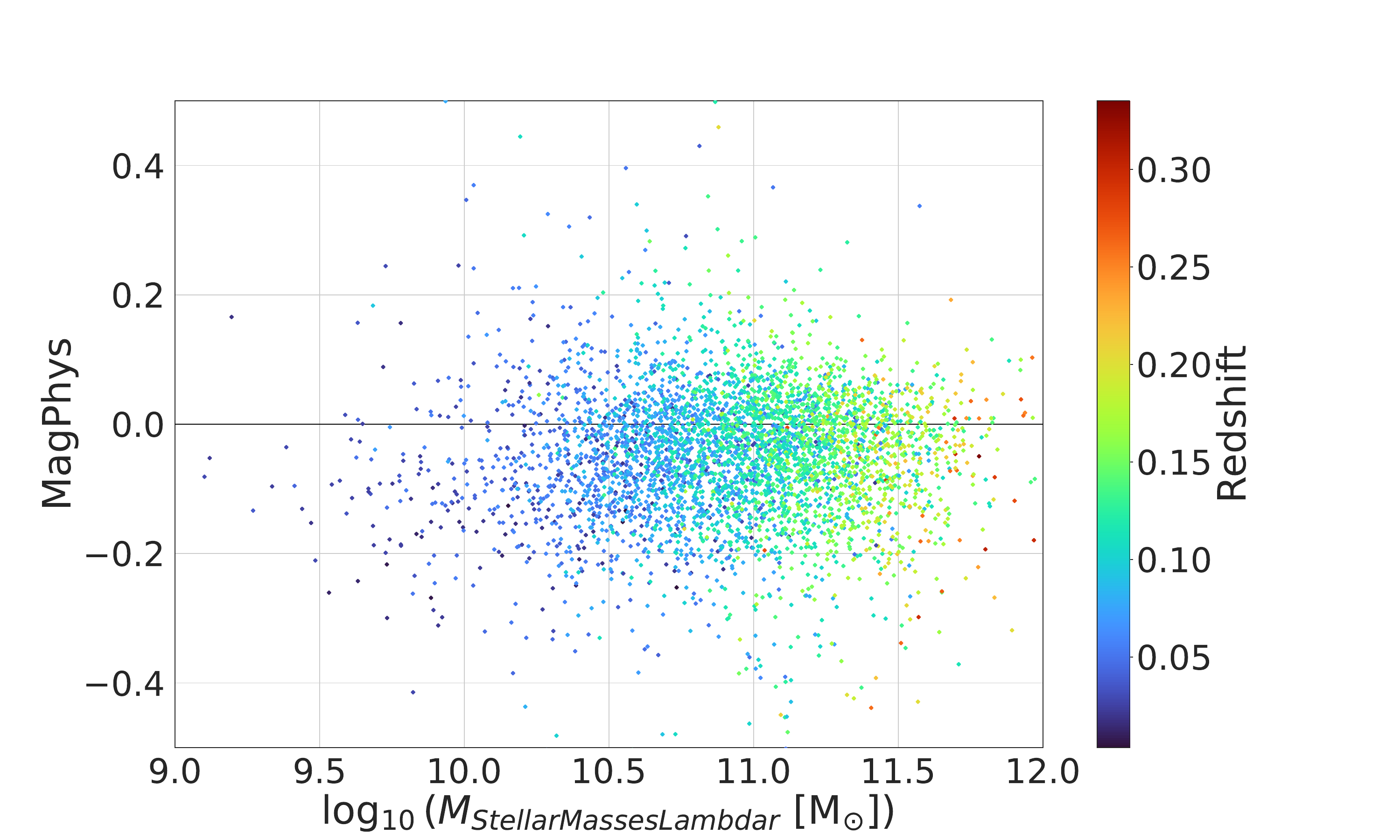} } 
\caption{Log differences between stellar masses from the three tables in the GAMA catalogue, plotted against the stellar masses from a specific GAMA table (indicated on the $x$-axis, with the other table is on the $y$-axis). The colour code shows the redshifts of the galaxies. Compare this with Fig.$~$\ref{fig:fraction}.}
\label{fig:fraction_GAMA}
\end{figure}

In this appendix, we examine the consistency of $M_\ast$ values across the GAMA catalogue, which serves as a reference for our findings in Section$~$\ref{sec:compare}. Fig.$~$\ref{fig:fraction_GAMA} shows the differences between stellar masses from various GAMA tables derived from SED fitting, plotted against the $\log_{10} M_\ast$ values from a specific table. These plots indicate that data points are closer to the 0 reference line, and the  redshift-dependent trends observed for the WISE-derived stellar masses in Fig.$~$\ref{fig:fraction} are not evident here. 
The differences between values from the StellarMasses and StellarMassesLambdar tables are minor, as these values are estimated using the same model but different photometry. Differences in the other plots are generally lower (between -0.5 and 0.5), although some are comparable to those in Fig.$~$\ref{fig:fraction}. $M_\ast$ derived from the \textsc{magphys} SED fitting code (MagPhys table) are generally lower than those from the \citet{Taylor_11} method (in the StellarMasses and StellarMassesLambdar tables).

\begin{figure}
\includegraphics[width = 3.3in]{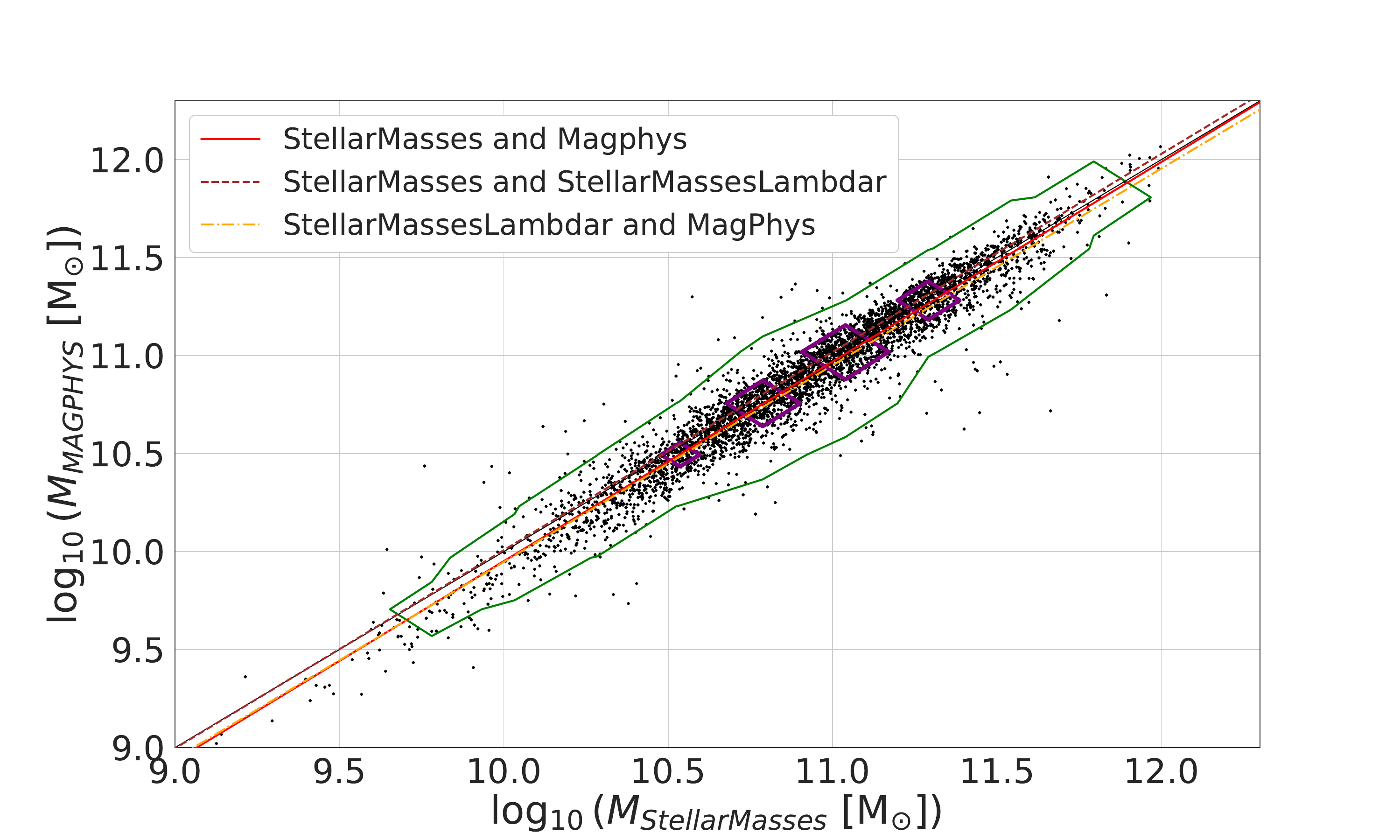}
\caption{Comparison of stellar masses in the GAMA catalogue. The figure shows the logarithms of stellar masses from the MagPhys table against the logarithms of values from StellarMasses table. The purple and green contours enclose 68 per cent and 95 per cent of galaxies, respectively. 
The black line marks the identity function. The red, purple, and orange lines represent linear relationships between stellar masses from the StellarMasses and MagPhys, StellarMasses and StellarMassesLambdar, and \text{StellarMassesLambdar} and MagPhys tables, respectively. In the legend, the first table corresponds to the x-axis, while the second to the y-axis. Compare this with Fig.$~$\ref{fig:compare}.}
\label{fig:GAMA}
\end{figure}

We compared $\log_{10} M_\ast$ values from different GAMA tables, similar to the comparison of tested methods and reference GAMA values shown in Fig.$~$\ref{fig:compare}. Fig.$~$\ref{fig:GAMA} compares values from the MagPhys and the StellarMasses tables, with the different lines representing the linear relationships between the tables indicated in the legend. Fitting parameters for both least squares and maximum likelihood methods are provided in Table$~$\ref{tab:GAMA_vals}. Our results show that the fitting parameters from both methods agree, with slopes close to 1 and standard deviations comparable to those reported in Table$~$\ref{tab:fitted_vals}.

\begin{table*}
\caption{Results of linear fitting ($y = a  x + b$) to the logarithms of stellar masses from different tables in the GAMA catalogue, using both maximum likelihood and least squares methods. For the maximum likelihood method, the standard deviation is listed in the $\sigma$ column. Note that errors in the data were not taken into account in this analysis.}
\label{tab:GAMA_vals}
\begin{tabular}{ccccccc}
\hline
\multicolumn{2}{c}{\textbf{Data}}           & \multicolumn{3}{c}{\textbf{Maximum likelihood fitting}} & \multicolumn{2}{c}{\textbf{Least square method}} \\
\textbf{x}           & \textbf{y}           & \textbf{a}     & \textbf{b}     & $\mathbf{\sigma}$     & \textbf{a}              & \textbf{b}             \\ \hline
StellarMasses        & StellarMassesLambdar & 1.011         & -0.102         & 0.0043                & 1.011$\pm$0.002         & -0.10$\pm$0.02         \\
StellarMasses        & MAGPHYS              & 1.018          & -0.231         & 0.1101                & 1.018$\pm$0.004         & -0.23$\pm$0.04        \\
StellarMassesLambdar & MAGPHYS              & 1.004          & -0.090          & 0.1076                & 1.004$\pm$0.004         & -0.09$\pm$0.04          \\ \hline
\end{tabular}
\end{table*}

\section{Host porbability as a function of stellar mass from Artale's table}\label{app:Artale_table}

\begin{figure}
    \centering
    \includegraphics[width=3.35 in]{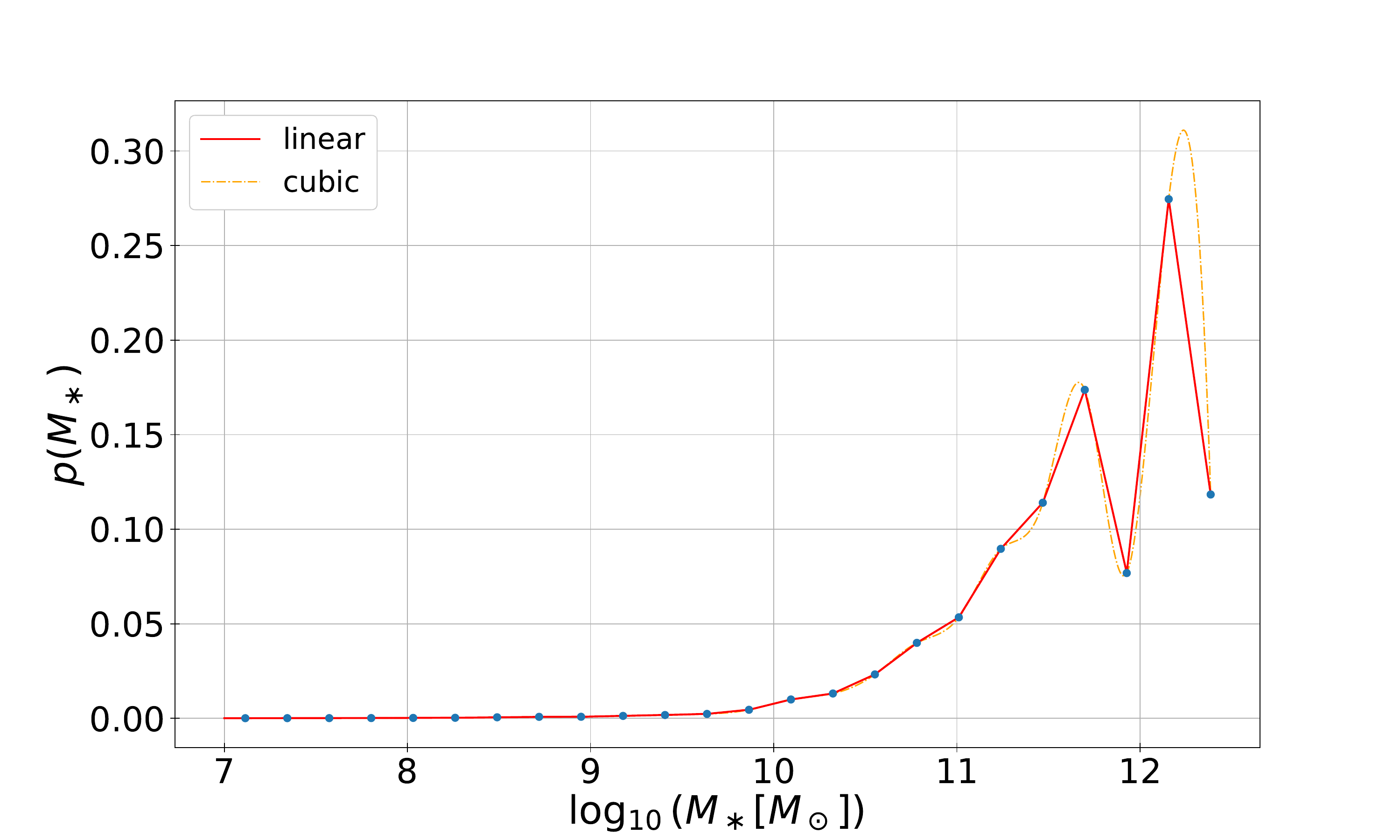}
    \caption{Host galaxy probability as a function of $\log_{10} M_\ast$ from \citet{Artale_2020} in the redshift range [0, 0.1]. The red line represents our linear interpolation, while the orange dotted line shows the cubic interpolation.}
    \label{fig:Artale_interpolated}
\end{figure}

We used the host galaxy probability–stellar mass relation for BBHs at $z \in [0, 0.1]$ from \citet{Artale_2020}. Fig.$~$\ref{fig:Artale_interpolated} shows the tabulated probability distribution as a function of $\log{10} M_\ast$. To assign probabilities to stellar masses not explicitly listed, we applied interpolation methods: linear (red) and cubic (orange, dotted). Given the limited data points and uncertainty in the fine structure, we chose linear interpolation for its robustness and computational simplicity.

\bsp	
\label{lastpage}
\end{document}